\documentclass{article}

 \usepackage[dandb, final]{neurips_2025}

\usepackage[utf8]{inputenc} %
\usepackage[T1]{fontenc}    %
\usepackage{hyperref}       %
\usepackage{url}            %
\usepackage{booktabs}       %
\usepackage{amsfonts}       %
\usepackage{nicefrac}       %
\usepackage{microtype}      %
\usepackage{xcolor}         %
\usepackage{multirow}
\usepackage{tabularx}
\usepackage{makecell}
\usepackage{pifont}
\usepackage{amsmath}
\usepackage{bm}
\usepackage[table]{xcolor}
\usepackage{graphicx}
\usepackage{caption}
\usepackage{subcaption}
\usepackage[inline]{enumitem}
\usepackage{array}
\usepackage{soul}
\newcommand{\ccmark}{\textcolor{green!60!black}{\ding{51}}}
\newcommand{\cxmark}{\textcolor{red}{\ding{55}}}

\definecolor{pastelgreen}{HTML}{DFF0D8}
\definecolor{pastelred}{HTML}{F2DEDE}
\definecolor{pastelorange}{HTML}{FCF8E3}

\title{Going with the Speed of Sound: Pushing Neural Surrogates into Highly-turbulent Transonic Regimes}

\author{%
  Fabian Paischer\thanks{Equal contribution} $\:^{1,2}$, Leo Cotteleer\footnotemark[1] $\:^{1}$, Yann Dreze\footnotemark[1] $\:^{1}$, Richard Kurle\footnotemark[1] $\:^{1}$, Dylan Rubini$\:^{1}$, \\ \textbf{Maurits Bleeker}$\:^{1}$,
  \textbf{Tobias Kronlachner}$\:^{1}$,
  \textbf{Johannes Brandstetter}$\:^{1,2}$ \\
  {$^1$~Emmi AI GmbH, Linz} \\
  {$^2$~ELLIS Unit, Institute for Machine Learning, JKU Linz}\\\\
}

\begin{document}

\maketitle

\begin{abstract}
The widespread use of neural surrogates in automotive aerodynamics, enabled by datasets such as DrivAerML and DrivAerNet++, has primarily focused on bluff-body flows with large wakes. 
Extending these methods to aerospace, particularly in the transonic regime, remains challenging due to the high level of non-linearity of compressible flows and 3D effects such as wingtip vortices. 
Existing aerospace datasets predominantly focus on 2D airfoils, neglecting these critical 3D phenomena.
To address this gap, we present a new dataset of CFD simulations for 3D wings in the transonic regime.
The dataset comprises volumetric and surface-level fields for around $30,000$ samples with unique geometry and inflow conditions. 
This allows computation of lift and drag coefficients, providing a foundation for data-driven aerodynamic optimization of the drag-lift Pareto front.
We evaluate several state-of-the-art neural surrogates on our dataset, including Transolver and AB-UPT, focusing on their out-of-distribution (OOD) generalization over geometry and inflow variations. 
AB-UPT demonstrates strong performance for transonic flowfields and reproduces physically consistent drag–lift Pareto fronts even for OOD wing configurations. 
Our results highlight its potential as an efficient and effective tool for rapid aerodynamic design exploration.
To facilitate future research, we open-source our dataset at  \url{https://huggingface.co/datasets/EmmiAI/Emmi-Wing}.
\end{abstract}

\section{Introduction}
Machine learning–based surrogates have recently emerged as powerful tools for accelerating aerodynamic design and analysis \citep{alkin2025abuptscalingneuralcfd,alkin_luminary_2025}. 
In automotive aerodynamics, large-scale datasets such as DrivAerML \citep{ashton2025drivaerml} and DrivAerNet++ \citep{elrefaie2024drivaernet,elrefaie2024drivaernetpp} have enabled neural models to predict complex bluff-body flows with remarkable accuracy, potentially reducing reliance on expensive Computational Fluid Dynamics (CFD) simulations.
In aerospace applications, the design goals and flow physics differ significantly from the automotive domain. 
Automobiles are bluff bodies with early flow separation, leading to high pressure drag \cite{kundu}. 
In contrast, aircrafts are streamlined bodies designed to maintain attached flow and optimize the lift-to-drag ratio \cite{Anderson2017}, which we refer to as drag–lift Pareto front.

While CFD is a fundamental design tool, its high computational cost creates a bottleneck in the design cycle. This has spurred the development of data-driven surrogate models \cite{alkin2024universal,alkin2025abuptscalingneuralcfd, wu2024Transolver, luo2025Transolver++,li2020fno,lu2021deeponet,li2023gino,li2023transformer,pfaff2021learning,li2020neural}, which learn the complex mapping between geometry and flow fields \cite{brunton2020Review}. 
However, extending these approaches to aerospace applications remains challenging. 
Transonic flight regimes involve intricate 3D phenomena, such as shock–boundary layer interactions and wingtip vortices, that are not captured with existing 2D airfoil datasets \citep{bonnet2023airfrans,kanchi_unifoil_2025,ramos_airfoilcfd_2023,schillaci_airfoilcfd_2021}. 
Moreover, the lack of publicly available high-fidelity 3D flow data limits the development and benchmarking of neural surrogates capable of generalizing across realistic aircraft geometries and operating conditions, as well as optimizing lift-to-drag performance.

To address these limitations, we introduce a new dataset of high-fidelity RANS simulations for 3D wings in the transonic regime.
To the best of our knowledge, this is the first publicly available dataset that comprehensively captures both geometric and inflow variations for realistic 3D configurations.
The dataset comprises approximately $30,000$ simulations with a unique combination of geometry and inflow parameters, providing volumetric as well as surface-level flow-field data.
These data enable the computation of aerodynamic performance metrics such as lift, drag, and drag–lift polars, thereby supporting data-driven design-space exploration and optimization.

Using our new dataset, we evaluate several state-of-the-art neural surrogates, including Transolver \citep{wu2024Transolver} and the recently proposed AB-UPT \citep{alkin2025abuptscalingneuralcfd}, with a focus on out-of-distribution generalization to unseen geometries and flow conditions.
Our results show that AB-UPT accurately predicts transonic flow fields and can reproduce physically consistent lift–drag trade-offs for unseen configurations, outperforming all competitors.
Overall, our study demonstrates that AB-UPT can approximate lift–drag Pareto fronts for unseen geometries, highlighting its potential as practical tool for rapid aerodynamic design exploration.%

\section{Methodology}
Our methodology centers on the generation of a high-fidelity dataset of 3D wings operating in the transonic regime, followed by the evaluation of state-of-the-art neural surrogate models.
Existing aerospace CFD datasets predominantly focus on subsonic 2D airfoils, as summarized in Table \ref{tab:dataset_comparison}, and therefore neglect critical 3D flow phenomena such as shock–boundary layer interactions and wingtip vortices.
We generate a new dataset comprising around $30,000$ cases with a unique combination of geometry and inflow parameters.
The data generation process is organized into three key components: 
\begin{enumerate*}[label=(\roman*)]
    \item the design of experiments for geometry and flow conditions,
    \item the setup and execution of high-fidelity CFD simulations, and
    \item the dataset split strategy used for training and evaluating neural surrogates.
\end{enumerate*}

\begin{table}[h]
    \centering
    \caption{Comparison of publicly available aerodynamic CFD datasets for machine learning.}
    \vspace{0.5em}
    \label{tab:dataset_comparison}
    \rowcolors{2}{gray!10}{white}
    \resizebox{\linewidth}{!}{
    \begin{tabular}{lccccl}
         \toprule
         \textbf{Dataset} & \textbf{Size} & \textbf{Dim.} & \textbf{Regime} & \textbf{Notes} \\
         \midrule
         AirfRANS \citep{bonnet2023airfrans} & $\sim$1\,000 & 2D & Subsonic & ML benchmark, varied AoA/Re \\
         UniFoil \citep{kanchi_unifoil_2025} & 500\,K & 2D &  Sub-/Transonic & Very large 2D dataset, wide Re/Mach range \\
         AirFoilCFD \citep{ramos_airfoilcfd_2023} & $\sim$18\,K & 2D &  Subsonic & 9k shapes, 2 AoA, fixed inflow \\
         AirFoilML \citep{schillaci_airfoilcfd_2021} & 2\,600 & 2D & Subsonic & NACA airfoils, fixed AoA/Re \\
         BlendedNet \citep{sung_blendednet_2025} & $\sim$10\,K & 3D &  Subsonic & Blended wing-body, 9 flight conds. \\
         \rowcolor{green!20} Emmi-Wing (Ours) & $\sim$ 30\,K & 3D & \textbf{Sub-/Transonic} & 30K unique geometries + inflows, drag/lift polars \\
         \bottomrule
    \end{tabular}}
\end{table}

\paragraph{CFD setup.}
We conduct simulations using the open-source CFD solver, OpenFOAM-v2506 \cite{openfoam}. The steady-state, compressible flow rhoSimpleFoam solver \cite{simpleAlgo} is used for all cases using the perfect gas assumption. 
For turbulence modeling, we selected the Spalart-Allmaras model which offers good computational efficiency and robust convergence for transonic wing flows with predominantly attached boundary layers, despite the presence of weak-to-moderate shocks~\cite{Alletto2024}.
Spatial discretization utilized second-order schemes including a van Leer limiter for momentum and a bounded upwind scheme for energy/pressure terms, with first-order upwind differencing applied to turbulence quantities for enhanced stability.
The wing surfaces are set with a no-slip condition, and freestream conditions are applied at the far-field boundaries. 
The angle of attack is imposed via the inflow boundary condition rather than by altering the wing geometry.
The meshing process relies on OpenFOAM's built-in snappyHexMesh to generate a high-quality, body-fitted mesh. Prismatic boundary layer meshing is used to achieve low $y^+$ values in the range $[50-200]$ in all cases. 
We verify our CFD setup using empirical data from wind tunnel measurements of the OneraM6 wing at different angles of attack \citep{Schmitt1979PressureDO,mani_oneram6_1997}.

\begin{figure}
    \centering
    \includegraphics[width=0.8\linewidth]{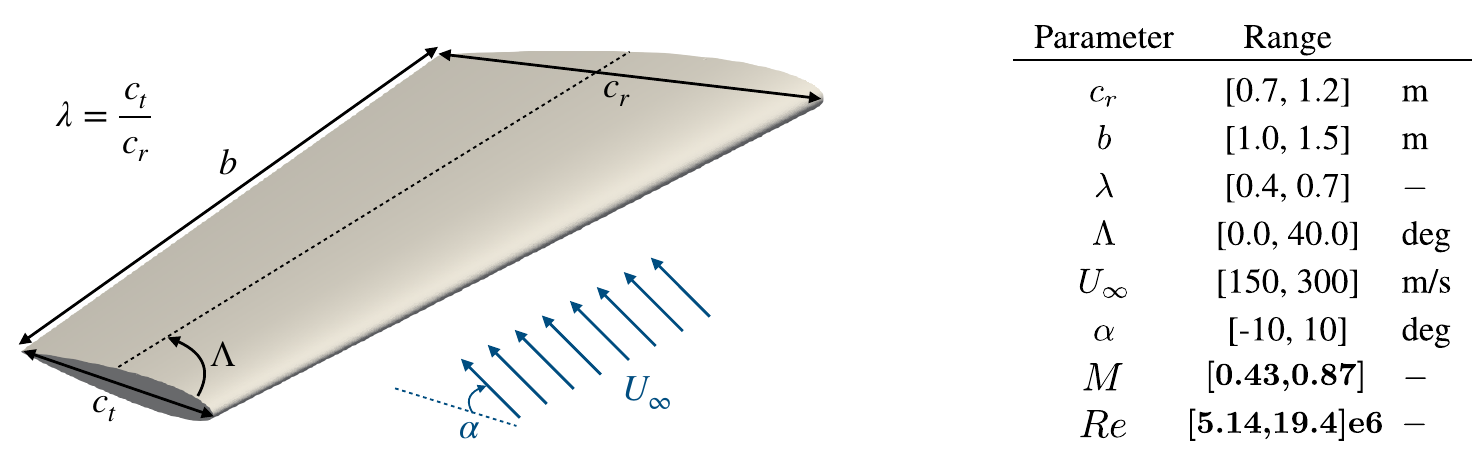}
    \caption{A visual representation of the parameterized 3D wing geometry alongside the four key geometric design parameters and the two inflow condition parameters with their respective sampling ranges used for the design of experiments. The geometry parameters include the span $(b)$, taper ratio $(\lambda)$, sweep angle $(\Lambda)$, and chord root $(c_r)$; inflow parameters are the inflow velocity $(U_{\infty})$ and angle of attack $(\alpha)$; dimensionless parameters are mach number ($M$) and Reynolds number ($Re$).}
    \label{fig:wing_scheme}
\end{figure}

\paragraph{Design of experiments.}
We select a 2D NACA0012 airfoil profile, which is extruded into 3D based on four geometric parameters: the span $(b)$, the taper ratio $(\lambda)$, the sweep angle $(\Lambda)$, and the root chord $(c_r)$. 
In addition, we vary two inflow parameters, namely velocity $(U_\infty)$ and angle of attack $(\alpha)$. 
To ensure a broad and diverse range of flow conditions, a total of $29,727$ unique simulation cases were created by randomly sampling each of the six variables from a uniform distribution within a application-relevant predefined range. The specific ranges for each parameter and their visual representation are depicted in Figure \ref{fig:wing_scheme}.

\begin{figure}
    \centering
    \includegraphics[width=\linewidth]{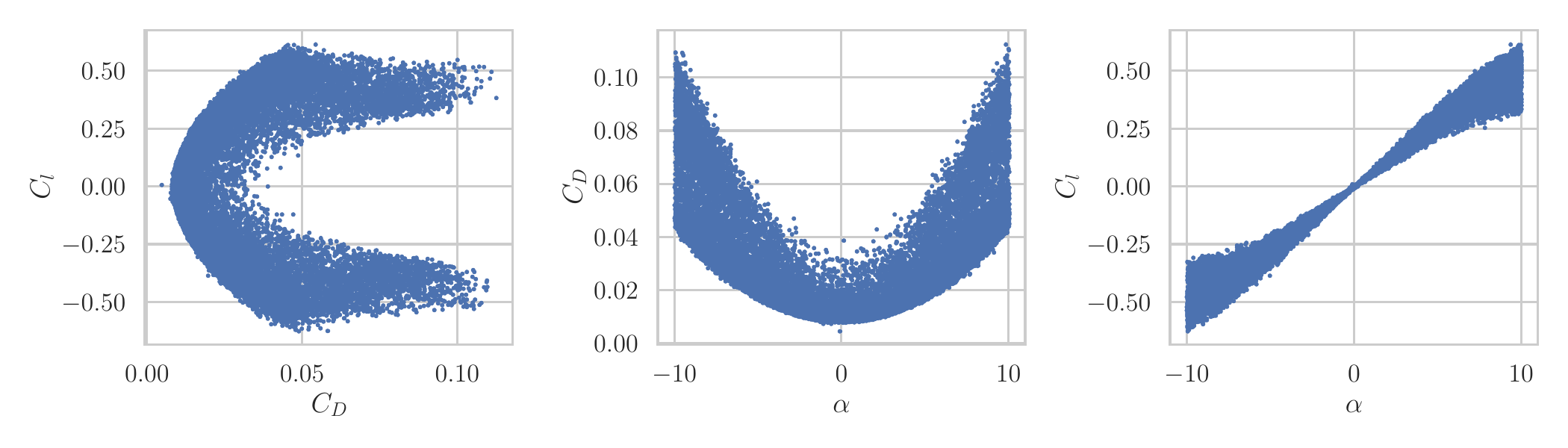}
    \caption{\textbf{Left:} Pareto frontier of drag $C_D$ versus lift $C_l$ coefficients for all cases in the dataset. \textbf{Middle:} $C_D$ as a function of angles of attack present in the dataset. \textbf{Right:} $C_l$ as a function of the angles of attack present in the dataset.}
    \label{fig:gt_drag_lift}
\end{figure}

The resulting dataset comprises $29,727$ cases, including both surface and volumetric data. The surface data consists of pressure $p_s$ and wall shear-stress $\boldsymbol{\tau}$ on the wing. The volumetric data contains the full 3D flow field, including pressure $p_v$, the velocity vector $\bm{u}=[u_x, u_y, u_z]$, its magnitude $||\bm{u}||$, and the vorticity vector $\boldsymbol{\omega} = [\omega_x,\omega_y,\omega_z]$ and its magnitude $||\boldsymbol{\omega}||$ at each point of the CFD mesh.
We provide visualizations of the lift-to-drag Pareto front as well as drag/lift coefficients over varying angles of attack for all cases in Figure \ref{fig:gt_drag_lift}.
In addition we visualize a set of 3D wing geometries for the minimum and maximum of each parameter in Figure \ref{fig:parameter_sweep_minmax} in the Appendix.
Finally, Figure \ref{fig:7910_sample_vis} shows a representative sample for a transonic sample of our dataset exhibiting flow detachment at high angles of attack and the formation of turbulent structures at the wingtip.

\begin{figure}
    \centering
    \includegraphics[width=.75\linewidth]{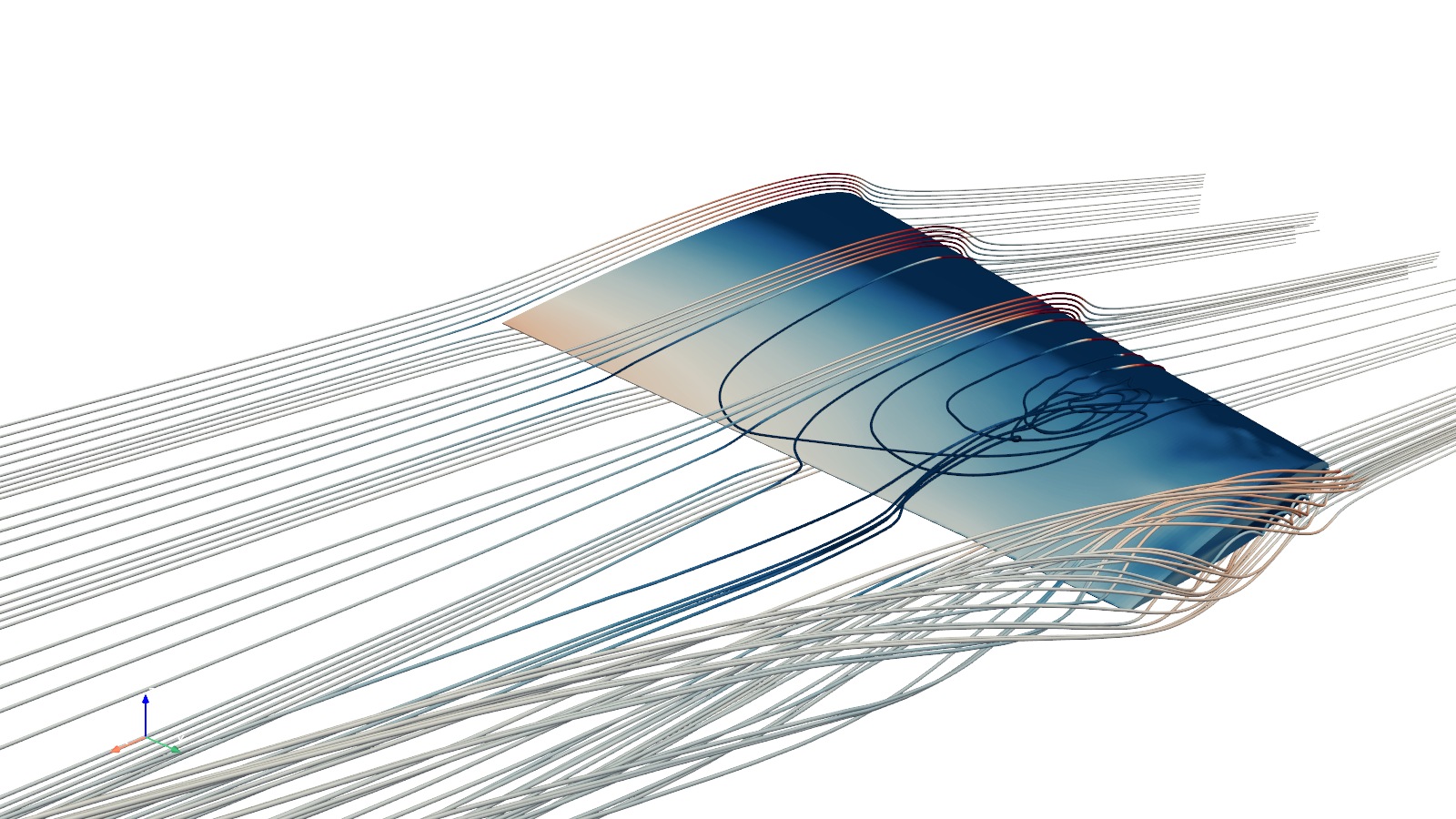}
    \caption{3D effects on wing.}
    \label{fig:7910_sample_vis}
\end{figure}

\paragraph{Dataset split.}
We partition the dataset into a training, a validation, and three test sets, including two in-distribution (ID) sets and one out-of-distribution (OOD) set.
The two ID sets are designed as \begin{enumerate*}[label=(\roman*)]
    \item \textbf{random selection}: a random selection of $1,000$ cases within the convex hull of the training set, and
    \item \textbf{interpolation}: a parameter region within the convex hull of the parameter space that spans $1,000$ unseen cases.
\end{enumerate*}
The former evaluates in-distribution performance, while the latter specifically targets the ability of the surrogate to interpolate between cases in an unseen ID parameter region.
The $1,000$ OOD cases are selected as the outermost points of the convex hull of the parameter space. 
To create the OOD set and the ID interpolation set, where we used an iterative convex hull peeling method and isolate the $1,000$ outermost data points and the $1,000$ innermost points, respectively. 
The remaining $25,727$ cases are used for training.

\textbf{Additional parameter scans.}
Practitioners are usually concerned with the optimization of the drag–lift Pareto front.
To this end, lift and drag coefficients are usually investigated on parameter scans of various angles of attack $\alpha$ and varying wing geometries to obtain the wing geometry resulting in the highest lift and lowest drag forces.
To asses the ability of neural surrogate models to capture these coefficients for different parameter scans we run additional cases with a unique geometry ($c_r=0.806$, $b=1.1963$, $\lambda=0.562$) that does not occur in the $29,727$ cases.
We run two parameter sweeps for this geometry, namely \begin{enumerate*}[label=(\roman*)]
    \item $\alpha \in \{ -30, -28, \ldots, 28, 30 \}$, and
    \item $\Lambda \in \{ 0, 10, 20, 30, 40, 50, 60, 70 \}$,
\end{enumerate*}
to evaluate in-distribution and OOD generalization. 
The latter sweep concerns a geometry parameter (sweep angle) as it is the geometry parameter apart from AoA with the most impact on lift/drag.
This results in another 248 cases solely used for evaluation.

\section{Experiments}
We conduct a series of experiments to evaluate the generalization capabilities of state-of-the-art neural surrogate models trained on our 3D transonic wing dataset.
Our analysis focuses on two main aspects: \begin{enumerate*}[label=(\roman*)]
    \item the in-distribution performance of different architectures on flowfield and aerodynamic coefficient prediction, and
    \item their out-of-distribution generalization to unseen geometries and inflow~conditions. 
\end{enumerate*}

\paragraph{Neural surrogates.}
We benchmark four neural surrogate models on our dataset: PointNet \cite{qi2017pointnet}, Transolver \cite{wu2024Transolver}, AB-UPT \cite{alkin2025abuptscalingneuralcfd}, and a Vision Transformer \cite{dosovitsky2021vit}. To incorporate inflow conditions and geometry design parameters, we add their embeddings to the input for PointNet, and for the transformer-based models we use a conditioning as in Diffusion Transformers \cite{Peebles2022ScalableDM}. 
All models are trained with a Mean Squared Error (MSE) loss applied to all fields for three different seeds.

\paragraph{Evaluation.}
We evaluate the predictive performance using the pointwise relative $\text{L}_\text{2}$ error, which is normalized by the $\text{L}_2$ norm of the ground truth.
In addition, we report the coefficient of determination ($R^2$) between drag/lift coefficients obtained from the best surrogate model with the ground truth coefficients.
Finally, we provide drag-lift Pareto fronts for parameter scans of the best performing surrogate model compared to ground truth to stress-test their generalization ability.

\paragraph{Results.}

\begin{table}[tb!]
\centering
\caption{Mean and standard deviation of relative L2 errors for surface fields ($p_s$, $\tau$) and volume fields ($p_v$, $\bm{u}$, $\boldsymbol{\omega}$) of different neural surrogates across three seeds.}
\label{tab:all_relative_errors}
\vspace{.5em}
\setlength{\tabcolsep}{6pt}
\resizebox{\linewidth}{!}{
\begin{tabular}{llccccc}
\toprule
\textbf{Test Set} & \textbf{Model} &
$p_s$ & $\tau$ & $p_v$ & $\bm{u}$ & $\boldsymbol{\omega}$ \\
\midrule

\multirow{4}{*}{Interpol}
& PointNet      & $0.052_{\pm 0.0005}$ & $0.227_{\pm 0.0005}$ & $0.050_{\pm 0.0000}$ & $0.152_{\pm 0.0010}$ & $0.238_{\pm 0.0050}$ \\
& Transformer   & 
$0.002_{\pm 0.0000}$ & $0.022_{\pm 0.0000}$ & $0.002_{\pm 0.0000}$ & $0.011_{\pm 0.0000}$ & $0.071_{\pm 0.0005}$ \\
& Transolver    & 
$0.002_{\pm 0.0000}$ & $0.022_{\pm 0.0000}$ & $0.002_{\pm 0.0000}$ & $0.011_{\pm 0.0000}$ & $0.090_{\pm 0.0022}$ \\
\rowcolor{gray!10} & \textbf{AB-UPT} & %
$0.002_{\pm 0.0000}$ & $0.022_{\pm 0.0009}$ & $0.002_{\pm 0.0005}$ & $0.010_{\pm 0.0008}$ & $0.064_{\pm 0.0043}$ \\
\midrule

\multirow{4}{*}{ID}
& PointNet      &
$0.101_{\pm 0.0000}$ & $0.441_{\pm 0.0020}$ & $0.096_{\pm 0.0000}$ & $0.299_{\pm 0.0005}$ & $0.374_{\pm 0.0015}$ \\
& Transformer   & 
$0.005_{\pm 0.0005}$ & $0.042_{\pm 0.0005}$ & $0.005_{\pm 0.0000}$ & $0.034_{\pm 0.0005}$ & $0.102_{\pm 0.0000}$ \\
& Transolver    &
$0.005_{\pm 0.0000}$ & $0.042_{\pm 0.0005}$ & $0.005_{\pm 0.0000}$ & $0.034_{\pm 0.0005}$ & $0.119_{\pm 0.0017}$ \\
\rowcolor{gray!10} & \textbf{AB-UPT} & 
$0.006_{\pm 0.0009}$ & $0.042_{\pm 0.0026}$ & $0.006_{\pm 0.0009}$ & $0.035_{\pm 0.0031}$ & $0.094_{\pm 0.0066}$ \\
\midrule

\multirow{4}{*}{OOD}
& PointNet      & 
$0.120_{\pm 0.0000}$ & $0.582_{\pm 0.0040}$ & $0.115_{\pm 0.0005}$ & $0.398_{\pm 0.0010}$ & $0.475_{\pm 0.0000}$ \\
& Transformer   & 
$0.008_{\pm 0.0000}$ & $0.057_{\pm 0.0009}$ & $0.007_{\pm 0.0005}$ & $0.051_{\pm 0.0005}$ & $0.125_{\pm 0.0005}$ \\
& Transolver    & 
$0.007_{\pm 0.0005}$ & $0.055_{\pm 0.0005}$ & $0.007_{\pm 0.0005}$ & $0.051_{\pm 0.0009}$ & $0.142_{\pm 0.0024}$ \\
\rowcolor{gray!10} & \textbf{AB-UPT} & 
$0.008_{\pm 0.0008}$ & $0.057_{\pm 0.0043}$ & $0.008_{\pm 0.0014}$ & $0.051_{\pm 0.0040}$ & $0.116_{\pm 0.0093}$ \\
\bottomrule
\end{tabular}}
\end{table}

We report results for volume and surface-level quantities in Table \ref{tab:all_relative_errors} for the different methods.
As expected we observe a consistent increase of error the more the model is pushed towards an OOD evaluation regime across all neural surrogate models.
Furthermore, our results show that except for PointNet, all surrogate approaches perform similarly on surface-level quantities.
However, on volume-level quantities that exhibit high variance, like vorticity, AB-UPT shows a significant improvement over competitors.
Therefore, we consider AB-UPT the strongest neural surrogate approach and evaluate it on the different parameter scans.

\begin{figure}[t!]
    \centering
    \begin{subfigure}{1.0\linewidth}
        \centering
        \includegraphics[width=\linewidth]{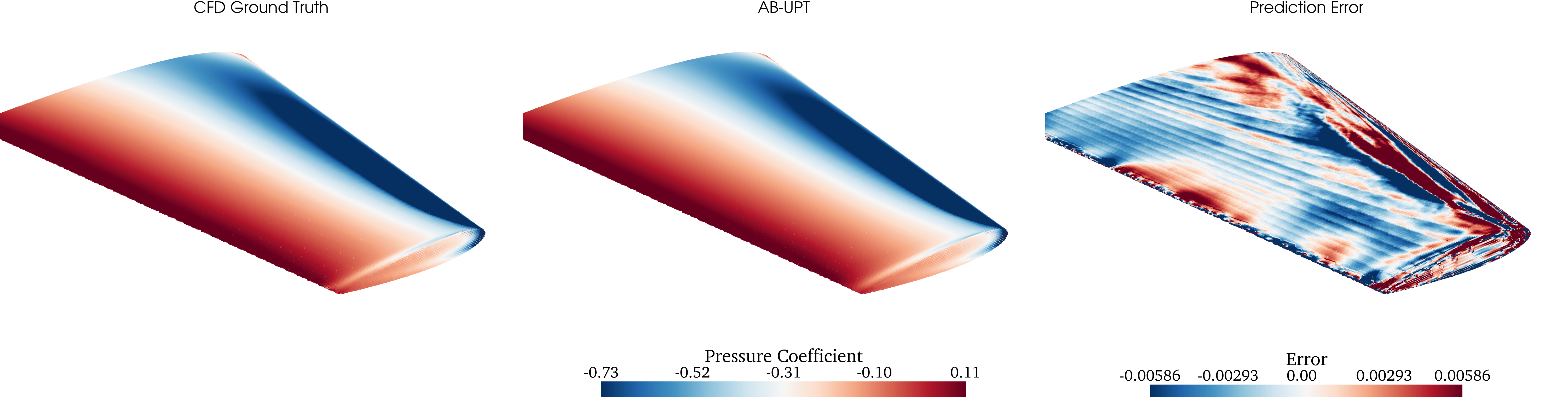}
        \caption{Pressure coefficient $(C_p)$}
        \label{fig:3D_comparison_test_extrapol_cp}
    \end{subfigure}
    \vspace{0.5cm}
    \begin{subfigure}{1.0\linewidth}
        \centering
        \includegraphics[width=\linewidth]{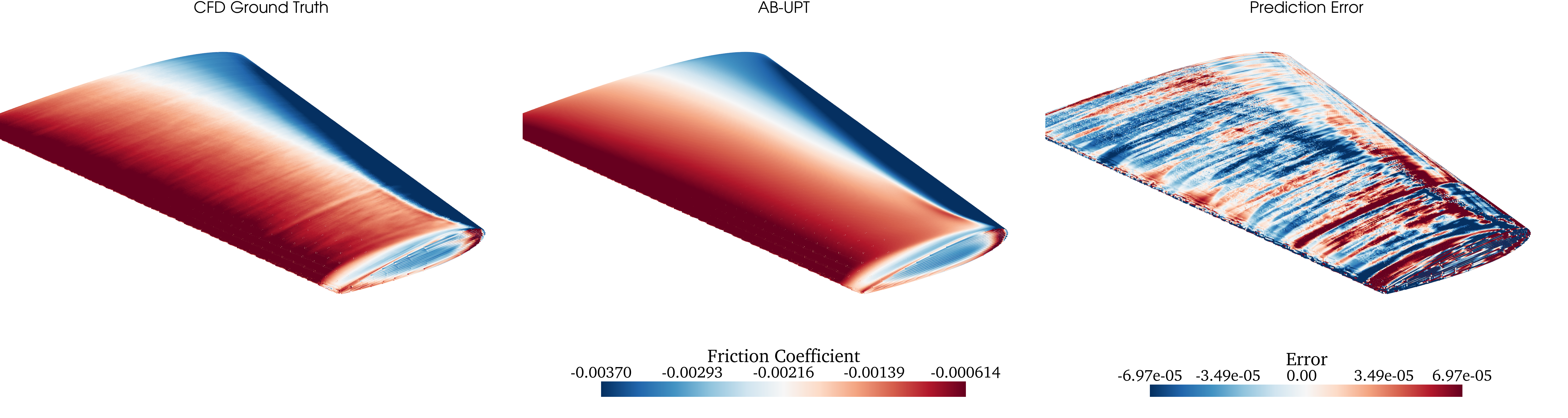}
        \caption{Friction coefficient $(C_f)$}
        \label{fig:3D_comparison_test_extrapol_cf}
    \end{subfigure}
    \caption{Comparison between surface field coefficients on the wing's surface of the CFD (left), AB-UPT surrogate (center) and the error between them (right). The case presented is from the the extrapolation test set with geometry and inflow conditioning parameters within the training range.}
    \label{fig:3D_comparison_test_extrapol_combined}
\end{figure}

We provide a qualitative analysis of the AB-UPT model on the OOD test set. 
Figure \ref{fig:profile_plot_test_extrapol} in the Appendix shows pressure and friction profiles of a randomly sampled test case of the OOD set at a normalized span location $y/b=0.75$. The corresponding 3D visualizations of the true surface fields, the predicted fields, and the prediction errors are shown in Figure~\ref{fig:3D_comparison_test_extrapol_combined}.
Remarkably, both pressure coefficient $(C_p)$ and friction coefficient $(C_f)$ match the ground truth closely.
Another interesting observation is that surface friction $(C_f)$ exhibits non-physical streaks on the wing surface.
Upon further investigation, we find that these artifacts are of numerical nature and non-physical.
Even though such artifacts may be present in the dataset,  AB-UPT seems not to be affected by them and still predicts a smooth surface friction field and the prediction error clearly shows that these artifacts are not captured. 
We attribute this to the inherent bias of neural networks towards low-frequency components \citep{teney2024neural} since the numerical artifacts mainly consist of high-frequency components.
Ultimately, this resembles another instance in which a neural surrogate potentially surpasses a numerical simulation due to inductive biases, first observed in \citep{koehler2025neural}.
In addition, it highlights the potential of AB-UPT acting as anomaly detector for data curation and removal of high-frequency numerical artifacts.

\begin{figure}[t!]
    \centering

    \begin{subfigure}{0.24\linewidth}
        \centering
        \includegraphics[width=\linewidth]{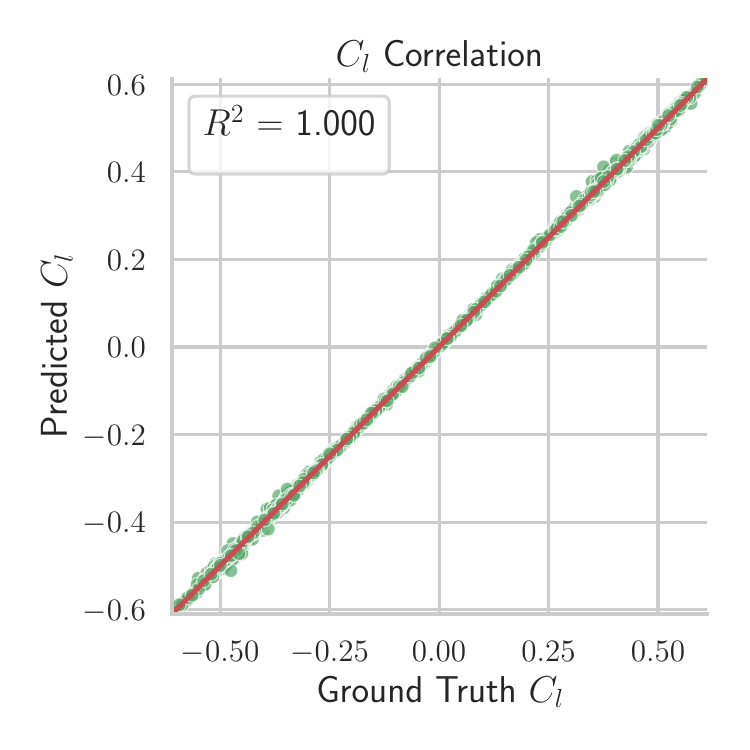}
        \caption{%
        }
        \label{fig:lift_corr_test_extrapol}
    \end{subfigure}
    \hfill
    \begin{subfigure}{0.24\linewidth}
        \centering
        \includegraphics[width=\linewidth]{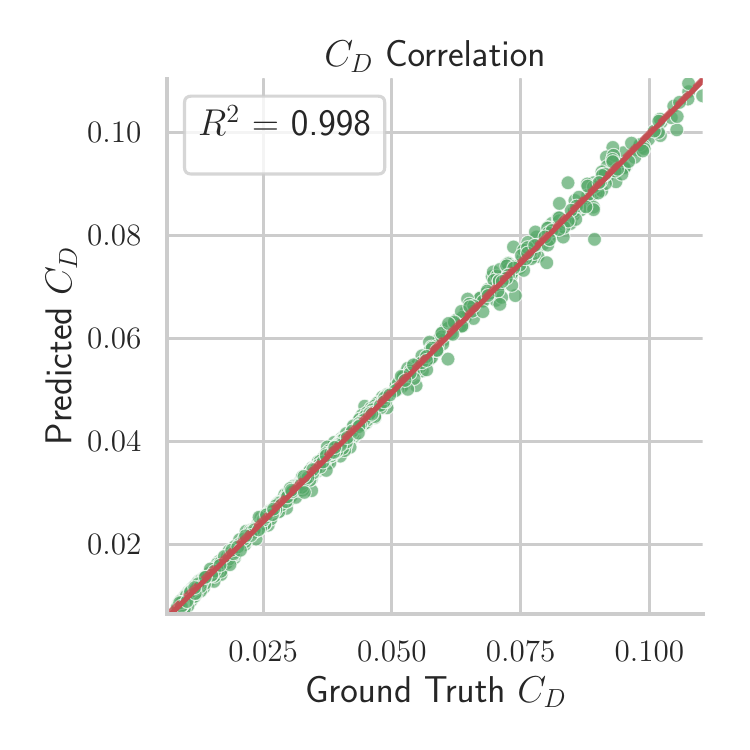}
        \caption{%
        }
        \label{fig:drag_corr_test_extrapol}
    \end{subfigure}
    \begin{subfigure}{0.24\linewidth}
        \centering
        \includegraphics[width=\linewidth]{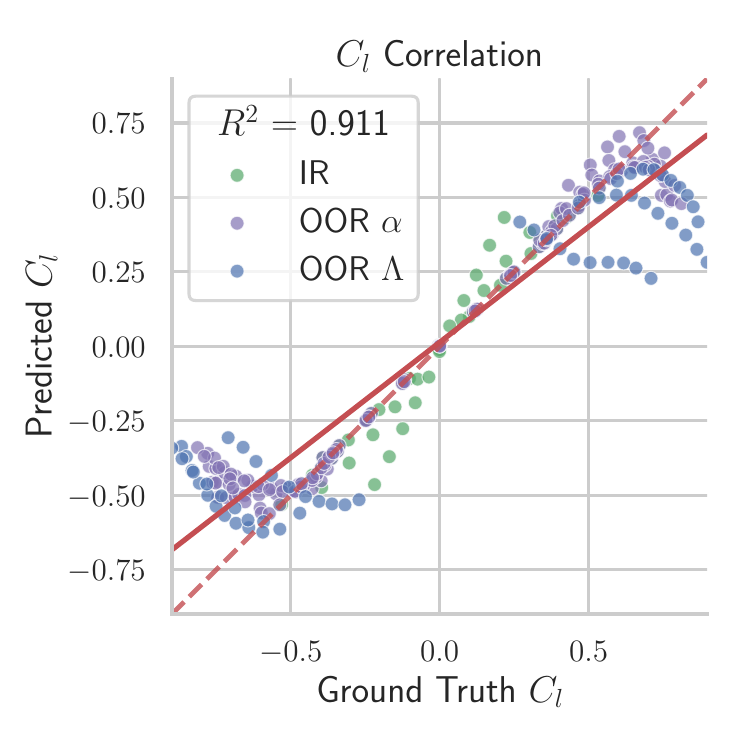}
        \caption{%
        }
        \label{fig:lift_correlation_scans}
    \end{subfigure}
    \hfill
    \begin{subfigure}{0.24\linewidth}
        \centering
        \includegraphics[width=\linewidth]{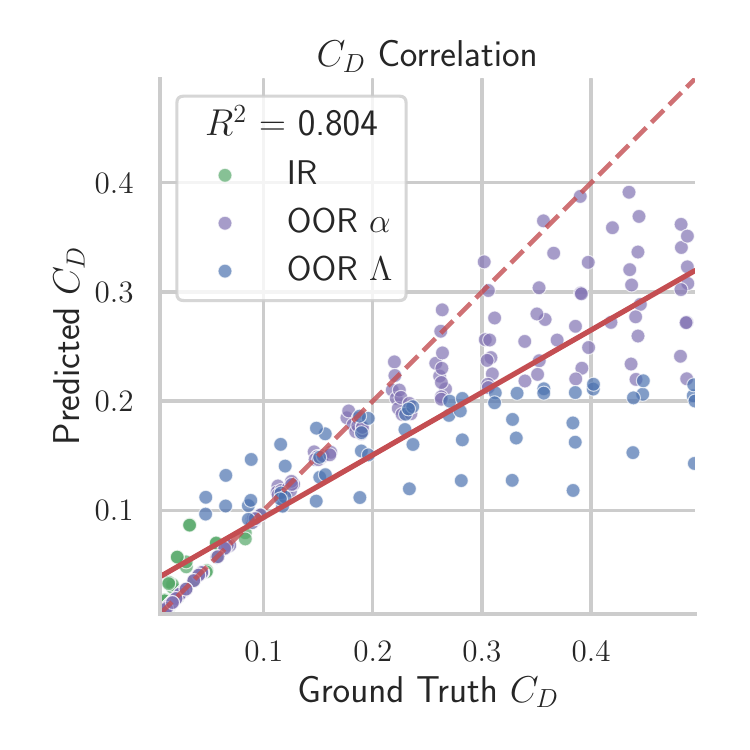}
        \caption{%
        }
        \label{fig:drag_correlation_scans}
    \end{subfigure}
    \caption{Correlation of predicted $C_l$ (\ref{fig:lift_corr_test_extrapol}) and $C_D$ (\ref{fig:drag_corr_test_extrapol}) from AB-UPT to the ground truth for all cases in the OOD test set. AB-UPT's predictions closely match the ground truth. Figures \ref{fig:lift_correlation_scans} and \ref{fig:drag_correlation_scans} depict correlation of $C_l$ and $C_D$ to ground-truth for all cases in the parameter scans with out-of-range parameters highlighted. AB-UPT still produces highly correlated results.}
    \label{fig:drag_lift_correlation_test_extrapol}
\end{figure}

In Figure \ref{fig:drag_lift_correlation_test_extrapol} we report correlation of drag $C_D$ and lift $C_l$ coefficients with the ground truth for all cases in the OOD test set.
The predictions of AB-UPT align closely with the ground-truth coefficients ($R^2=1.0$ for $C_l$ and $R^2=0.998$ for $C_D$).
Furthermore we report the error distribution across cases in the OOD test set for different combinations of drag and lift in Figure \ref{fig:average_l2err_test_extrapol}.
Intriguingly, we find that AB-UPT is most error-prone in high $C_D$ regimes, which are dominated by errors in $\tau$ and rather far from the tangent on the lift-drag Pareto front.
This highlights that AB-UPT is a promising contender for wing design optimization.

\paragraph{Parameter scans.}
We evaluate AB-UPT on the parameter scans that comprise 248 simulations with sweeps over $\Lambda$ and $\alpha$ with the remaining parameters fixed.
We report correlation of the predicted values for $C_l$ and $C_D$ in Figure \ref{fig:lift_correlation_scans} and \ref{fig:drag_correlation_scans}, respectively, with highlighted cases for out-of-range (OOR) values of $\alpha$ and $\Lambda$.
Remarkably, AB-UPT maintains high correlation with the ground-truth ($R^2=0.911$ for $C_l$, $R^2=0.804$ for $C_D$) even when observing values that are far beyond the range of values for $\alpha$ or $\Lambda$ it has been trained on.
We complement this observation with relative errors for surface quantities $p_s$ and $\tau$, as well as an average thereof in Figure \ref{fig:errs_sweep_vs_aoa}.
Interestingly, AB-UPT seems to perform better for positive angles of attack than for negative ones, even though the dataset was sampled uniformly between them.

\begin{figure}[t!]
    \centering

    \includegraphics[width=0.4\linewidth]{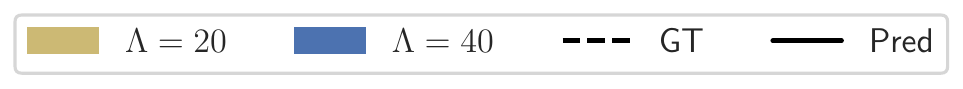}
    
    \begin{subfigure}{0.32\linewidth}
        \centering
        \includegraphics[width=\linewidth]{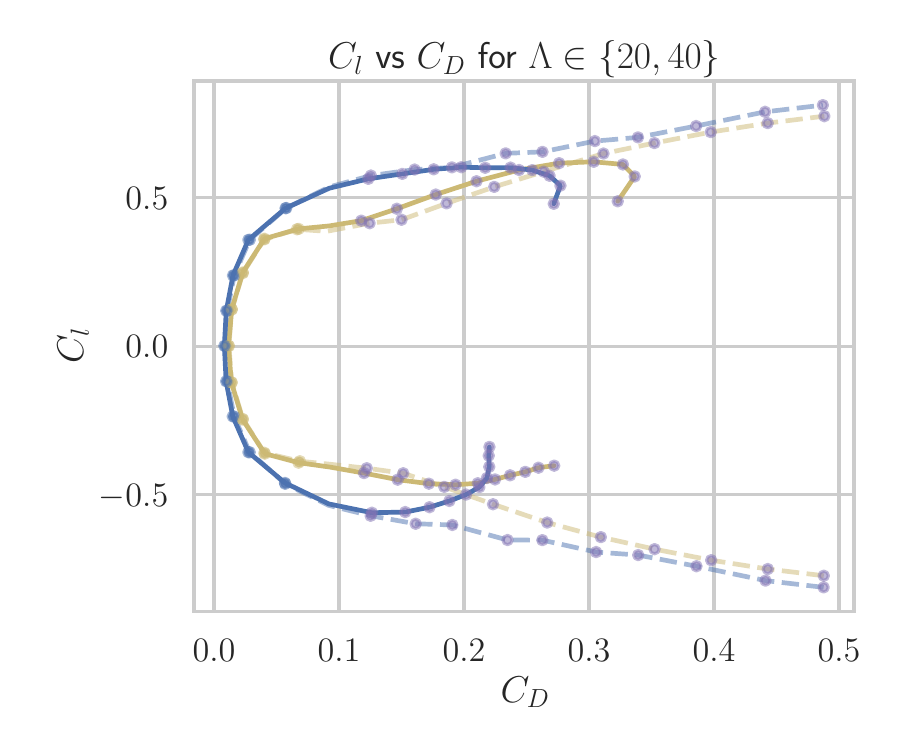}
    \end{subfigure}
    \hfill
    \begin{subfigure}{0.32\linewidth}
        \centering
        \raisebox{.3cm}{\includegraphics[width=\linewidth]{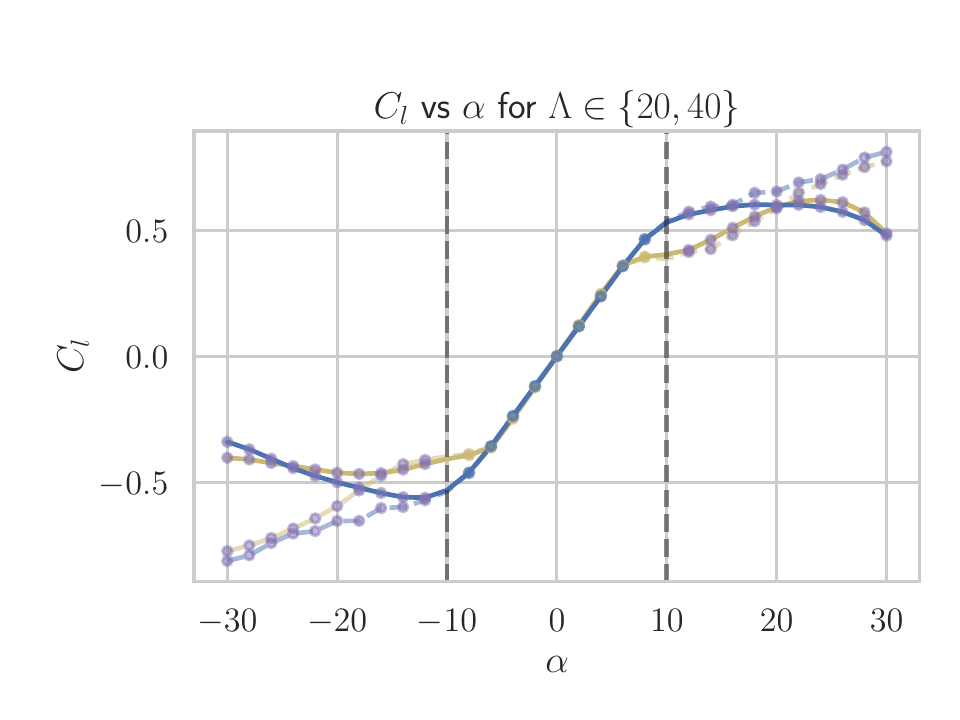}}
    \end{subfigure}
    \hfill
    \begin{subfigure}{0.32\linewidth}
        \centering
        \raisebox{.3cm}{\includegraphics[width=\linewidth]{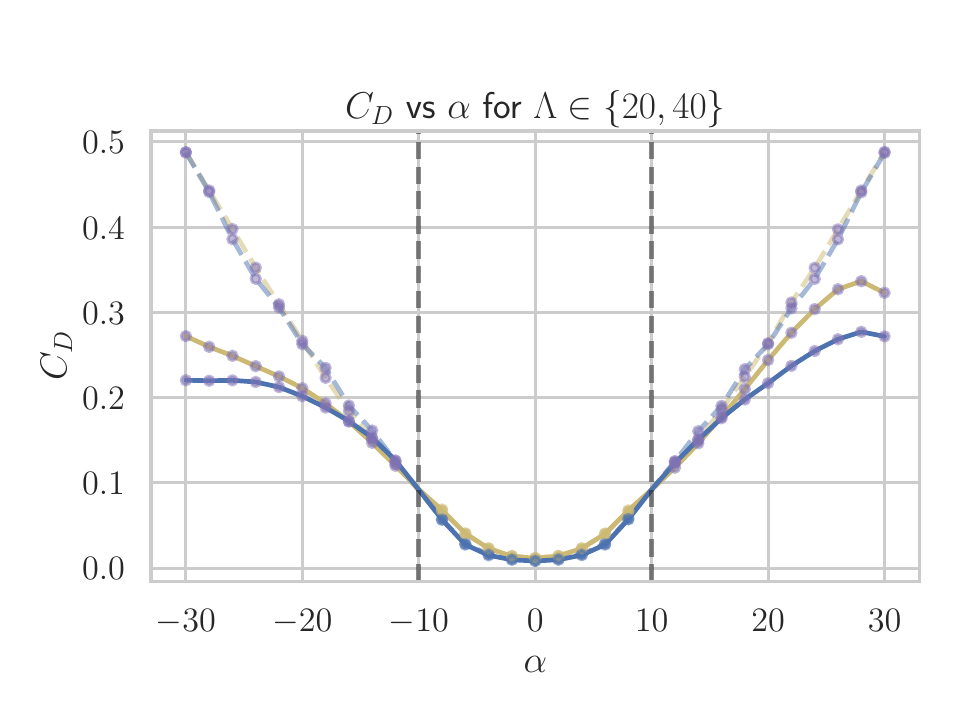}}
    \end{subfigure}
    \caption{Ground-truth vs predicted $C_l$ and $C_D$ for parameter scans over $\alpha \in \{-30,\ldots,30\}$ and $\Lambda \in \{ 20,40 \}$. Out-of-range cases are highlighted by purple markers, and the ground-truth is represented as dashed line. AB-UPT reproduces the drag-lift Pareto front well (left), but introduces error in high regimes of $\alpha > 20$ for both $C_l$ (middle) and $C_D$ (right).}
    \label{fig:gt_vs_pred_scans_sweep_20_40}
\end{figure}

\begin{figure}[t!]
    \centering

    \includegraphics[width=0.4\linewidth]{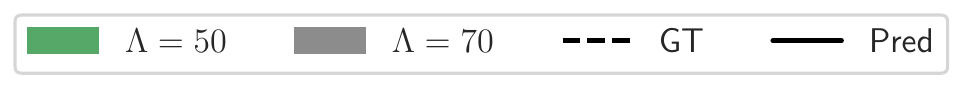}
    
    \begin{subfigure}{0.32\linewidth}
        \centering
        \includegraphics[width=\linewidth]{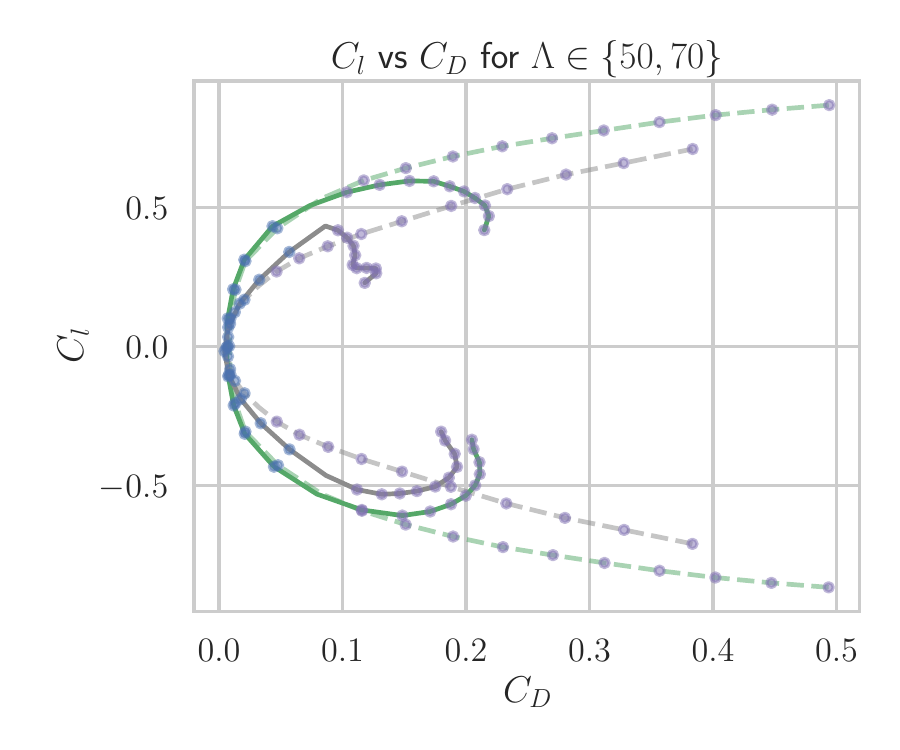}
    \end{subfigure}
    \hfill
    \begin{subfigure}{0.32\linewidth}
        \centering
        \raisebox{.3cm}{\includegraphics[width=\linewidth]{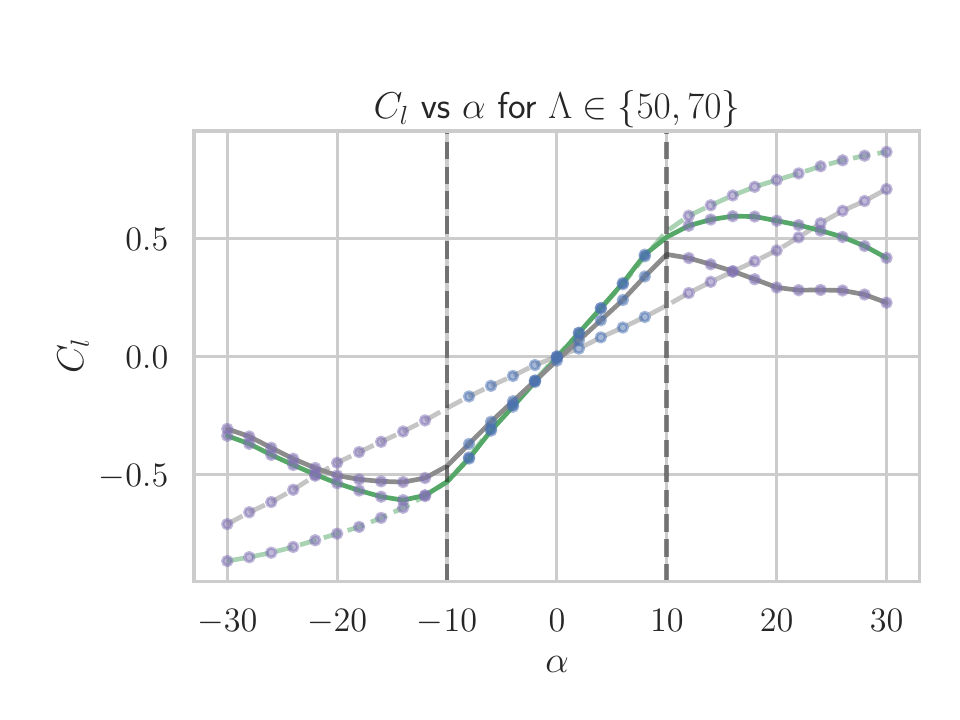}}
    \end{subfigure}
    \hfill
    \begin{subfigure}{0.32\linewidth}
        \centering
        \raisebox{.3cm}{\includegraphics[width=\linewidth]{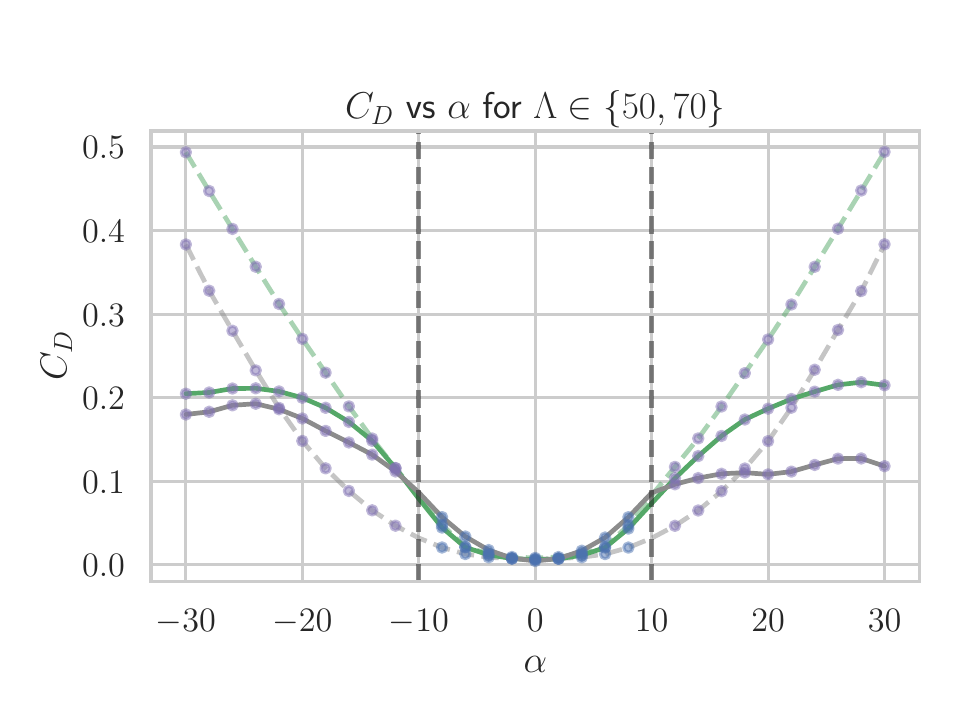}}
    \end{subfigure}
    \caption{Ground-truth vs predicted $C_l$ and $C_D$ for parameter scans over $\alpha \in \{-30,\ldots,30\}$ and $\Lambda \in \{50,70\}$. Out-of-range cases are highlighted by purple markers, and the ground-truth is represented as dashed line. AB-UPT reproduces the drag-lift Pareto front for the most part (left), but error accumulates especially for higher values of $\alpha$ and $\Lambda = 70$ for both $C_l$ (middle) and $C_D$ (right).}
    \label{fig:gt_vs_pred_scans_sweep_50_70}
\end{figure}

In addition, we report the predicted drag-lift Pareto fronts for a selected set of sweep angles ($\Lambda \in \{ 20,40,50,70\}$) and compare them to the ground-truth in Figures \ref{fig:gt_vs_pred_scans_sweep_20_40} and \ref{fig:gt_vs_pred_scans_sweep_50_70}. 
Intriguingly, we observe minor deviations from the ground truth for values up to $\alpha \sim 20^\circ$, which is far beyond the range the model has been trained on.
Furthermore, the tangent on the drag-lift Pareto front is well captured up to $\Lambda = 50$, which is again out-of-range compared to the training set. 
For larger $\Lambda > 50$ we observe larger divergence from the ground-truth.
Finally, we report pressure coefficient $C_p$ and friction coefficient $C_f$ at a normalized span length of $y/b=0.5$ for a test case with out-of-range $\alpha=20^\circ$ and sweep angle $\Lambda = 40$.
We also show $C_p$ and $C_f$ for the full 3D wing in Figures \ref{fig:3D_comparison_test_scans_cp} and \ref{fig:3D_comparison_test_scans_cf}, respectively.

\paragraph{Design Optimization.}
There are two avenues of performing design optimization, namely 
\begin{enumerate*}[label=(\roman*)]
    \item CFD-based optimization \citep{martins2020perspectives,lyu_aerodynamic_2015,he_robust_2019}, and
    \item data-driven surrogate-based optimization \citep{li_data-based_2021,li_efficient_2020,bartoli_adaptive_2019,benaouali_multidisciplinary_2019}.
\end{enumerate*}
The former typically employs adjoint methods \citep{kenway_effective_2019,shi_natural_2020} combined with gradient-based optimizers, requiring the repeated solution of the governing PDE system.
Traditional surrogate-based approaches instead usually construct low-dimensional response surfaces to predict quantities of interest directly from design parameters, significantly reducing computational cost but limiting physical fidelity.
Our AB-UPT directly operates on the physical fidelity of the CFD simulation, while bypassing the need to solve the governing PDE system.

We utilize our dataset to demonstrate how AB-UPT can be used for rapid design exploration and optimization. 
This requires real-time generation of new geometries from the design parameters in a suitable format that can be processed by the model, since the optimization explores new designs. 
Furthermore, gradient-based design optimization methods require the geometry generation to be differentiable. 
Relying on the traditional simulation meshes (e.g., taking cell centers as inputs) prohibits this workflow, as the meshing process is computationally intensive and generally not differentiable.

To address this problem, we train AB-UPT to take a computer-aided-design (CAD) representation as input to the geometry encoder and the anchor tokens (cf.~Section 4.6 in \cite{alkin2025abuptscalingneuralcfd}). 
Specifically, we generate stereolithography (STL) data in a fully differentiable manner and at arbitrary resolution by lofting a parametric NACA0012 profile along the wing span, applying affine transformations defined by the root chord, taper ratio, and sweep angle. 
This enables the generation of meshes for arbitrary wing configurations in milliseconds, enabling continuous design exploration. 

The cell centers of a low resolution STL mesh serve as input to the AB-UPT geometry encoder and surface anchor tokens, while cell centers from a higher resolution mesh serve as query tokens to predict higher-fidelity surface fields from which we can derive aerodynamic coefficients such as lift and drag.
We compute the drag and lift coefficients according to Eq.~\ref{eq:drag_and_lift_coefficient} for a particular mesh and use the lift-to-drag ratio $\varepsilon = \nicefrac{C_D}{C_l}$ as a measure of aerodynamic efficiency. 
Finally, we formulate the optimization problem as $\max_{\boldsymbol{\phi}} \varepsilon$, where the design vector $\boldsymbol{\phi} = \{ c_r,b,\lambda,\Lambda,U_\infty,\alpha \}$ comprises the four geometry parameters and two inflow conditions.
Note that while we consider optimizing both geometry and inflow conditions, it is straightforward to define a constrained optimization problem such as optimizing the geometry for a specific cruise flight speed. 

We explore three optimization methods,
\begin{enumerate*}[label=(\roman*)]
    \item gradient-based optimization (Adam \cite{kingma2014adam}),
    \item evolutionary strategies (CMA-ES \cite{Hansen2001CMAES, Hansen2023cmaes}), and
    \item Bayesian optimization (Gaussian Process \cite{Jones1998EfficientGO}).
\end{enumerate*}
We initialize each method with parameter values in the center of the training data range and bound the search space to the training range. 
Each method is run for 2 minutes on an H100 GPU until convergence and the resulting parameters, as well as the corresponding aerodynamic coefficients, are summarized in Table~\ref{tab:opt_methods}. 
Notably, the gradient-based method converges to a different solution than the evolutionary strategy and Bayesian optimization, with a distinct optimal flight speed, sweep angle, and angle of attack.
We hypothesize that this behavior is due to the gradient-based approach being more prone to local optima. 
To validate the newly found solution, we identify its nearest neighbor in the training set and compare the corresponding lift-to-drag ratio.
Indeed, we observe that the nearest neighbors in parameter space align well with the newly explored solutions.
Remarkably, though, the explored solutions exhibit slightly higher lift-to-drag ratio, indicating that our design optimization loop found more efficient 3D wing configurations than present in the dataset it has been trained on.
These results demonstrate the promise of leveraging neural surrogates for aerodynamic design optimization.

\begin{table}[h]
\centering
\caption{Performance comparison of optimization methods, running each method for approximately 2 minutes. %
The model forward and backward (in case of gradient-based) is computed on an H100 GPU in float16 precision. 
The gradient-based methods finds a different optimum compared to the gradient-free methods with similar lift-to-drag ratio ($\varepsilon$) %
We compare the optima to the nearest neighbor in the dataset (NN) %
and also show the best parameter set in the dataset, which is the nearest neighbor of the solution found by the gradient-based method. 
}
\vspace{0.5em}
\label{tab:opt_methods}
\renewcommand{\arraystretch}{1.2}
\resizebox{\textwidth}{!}{
\begin{tabular}{l c c c c c c c c c c}
\toprule
& & \multicolumn{6}{c}{\textbf{Input parameters}} & \multicolumn{3}{c}{\textbf{Coefficients}} \\
\cmidrule(lr){3-8} \cmidrule(lr){9-11}
\textbf{Method} & \# steps & $c_r\;[\text{m}]$ & $b\;[\text{m}]$ & $\lambda$ & $\Lambda\;[^\circ]$ & $U_\infty\;[\text{m/s}]$ & $\alpha\;[^\circ]$ & $C_d$ & $C_l$ & $\varepsilon$ \\
\midrule
Gradient    &  900  & 0.704 & 1.499 & 0.401 & 33.03 & 150.1 & 5.060 & 0.0179 & 0.3281 & 18.36 \\
Evolutionary &  2700 & 0.701 & 1.498 & 0.401 & 39.96 & 209.0 & 4.746 & 0.0165 & 0.3040 & 18.43 \\
Bayesian  &  100  & 0.700 & 1.500 & 0.400 & 40.00 & 212.4 & 4.664 & 0.0163 & 0.3002 & 18.43 \\
\midrule
Gradient (NN) & - & 0.710 & 1.468 & 0.443 & 27.72 & 154.3 & 4.371 & 0.0160 & 0.2905 & 18.12 \\
Evolutionary (NN) & - & 0.750 & 1.487 & 0.435 & 39.67 & 232.9 & 3.723 & 0.0138 & 0.2395 & 17.34 \\
Bayesian (NN) & - & 0.750 & 1.487 & 0.435 & 39.67 & 232.9 & 3.723 & 0.0138 & 0.2395 & 17.34 \\
Best $\varepsilon$ in dataset & - & 0.710 & 1.468 & 0.443 & 27.72 & 154.3 & 4.371 & 0.0160 & 0.2905 & 18.12 \\
\bottomrule
\end{tabular}}
\end{table}

\section{Discussion and Limitations}
\label{sec:discussion}

To the best of our knowledge we are the first to introduce a dataset of 3D wing geometries in the transonic regime and to show that neural surrogates can be used for aerodynamic design optimization.
However, there are some shortcomings to our work, mainly related to the dataset generation process.
Generally, we emphasize that there are many knobs to turn with respect to data generation that can lead to uncertainty in the resulting simulations, including mesh quality, turbulence-model assumptions, convergence criteria, and, importantly, solver fidelity.
These effects are exacerbated when moving from the subsonic to the transonic regime, where severe shockwaves and turbulence occur.
In this work we relied on OpenFOAM, as it is advantageous for accessibility and automation.
Furthermore, our current study relies on relatively simple wing geometries (NACA0012-based) and steady-state RANS simulations to reduce the cost of data generation.
As a result, inherent unsteady phenomena are not captured by the solver.
Despite these factors, our results demonstrate that neural surrogates trained on our dataset retain significant practical value. 
They enable efficient exploration of drag–lift Pareto fronts and exhibit potential for anomaly detection. 

\section{Conclusion}

We present a first comprehensive dataset of high-fidelity RANS simulations for 3D wings operating in the transonic regime.
Each simulation case comes with unique geometry and inflow parameters and provides volume and surface-level flowfields.
We evaluate state-of-the-art neural surrogates on our dataset and find that AB-UPT outperforms other state-of-the-art and provides strong performance even for unseen cases.
Furthermore, we demonstrate that AB-UPT accurately reproduces drag-lift Pareto fronts even for cases with parameters far beyond its training regime, establishing itself as a contender for real-time wing geometry optimization.

In the future our aim is to investigate the use of neural surrogates for anomaly detection to provide an iterative data generation pipeline and ensure data curation.
Furthermore, we plan to explore other numerical solvers and compare to empirical data if available.
Finally, we aim to push the boundary of efficiency such that our surrogates can be used as practical tool to explore the aerodynamic design space and optimize the lift-to-drag performance of unseen geometries in real time.

\bibliographystyle{unsrt}
\bibliography{references}

@article{lu2021deeponet,
   title={Learning nonlinear operators via DeepONet based on the universal approximation theorem of operators},
   volume={3},
   ISSN={2522-5839},
   DOI={10.1038/s42256-021-00302-5},
   number={3},
   journal={Nature Machine Intelligence},
   publisher={Springer Science and Business Media LLC},
   author={Lu, Lu and Jin, Pengzhan and Pang, Guofei and Zhang, Zhongqiang and Karniadakis, George Em},
   year={2021},
   month=mar, pages={218–229} }

@article{li2020fno,
  author       = {Zongyi Li and
                  Nikola B. Kovachki and
                  Kamyar Azizzadenesheli and
                  Burigede Liu and
                  Kaushik Bhattacharya and
                  Andrew M. Stuart and
                  Anima Anandkumar},
  title        = {Fourier Neural Operator for Parametric Partial Differential Equations},
  journal      = {CoRR},
  volume       = {abs/2010.08895},
  year         = {2020},
  url          = {https://arxiv.org/abs/2010.08895},
  eprinttype    = {arXiv},
  eprint       = {2010.08895},
  bibsource    = {dblp computer science bibliography, https://dblp.org}
}

@misc{li2023gino,
      title={Geometry-Informed Neural Operator for Large-Scale 3D PDEs}, 
      author={Zongyi Li and Nikola Borislavov Kovachki and Chris Choy and Boyi Li and Jean Kossaifi and Shourya Prakash Otta and Mohammad Amin Nabian and Maximilian Stadler and Christian Hundt and Kamyar Azizzadenesheli and Anima Anandkumar},
      year={2023},
      eprint={2309.00583},
      archivePrefix={arXiv},
      primaryClass={cs.LG},
}

@inproceedings{elrefaie2024drivaernetpp,
 author = {Elrefaie, Mohamed and Morar, Florin and Dai, Angela and Ahmed, Faez},
 booktitle = {Advances in Neural Information Processing Systems},
 editor = {A. Globerson and L. Mackey and D. Belgrave and A. Fan and U. Paquet and J. Tomczak and C. Zhang},
 pages = {499--536},
 publisher = {Curran Associates, Inc.},
 title = {DrivAerNet++: A Large-Scale Multimodal Car Dataset with Computational Fluid Dynamics Simulations and Deep Learning Benchmarks},
 url = {https://proceedings.neurips.cc/paper_files/paper/2024/file/013cf29a9e68e4411d0593040a8a1eb3-Paper-Datasets_and_Benchmarks_Track.pdf},
 volume = {37},
 year = {2024}
}

@inproceedings{elrefaie2024drivaernet,
author = {Elrefaie, Mohamed and Dai, Angela and Ahmed, Faez},
title = {DrivAerNet: A Parametric Car Dataset for Data-Driven Aerodynamic Design and Graph-Based Drag Prediction},
volume = {Volume 3A: 50th Design Automation Conference (DAC)},
series = {International Design Engineering Technical Conferences and Computers and Information in Engineering Conference},
publisher = {Curran Associates, Inc.},
pages = {V03AT03A019},
year = {2024},
month = {08}
}

@inproceedings{wu2024Transolver,
  title={Transolver: A Fast Transformer Solver for PDEs on General Geometries},
  author={Haixu Wu and Huakun Luo and Haowen Wang and Jianmin Wang and Mingsheng Long},
  booktitle={International Conference on Machine Learning},
  year={2024}
}

@misc{luo2025Transolver++,
      title={Transolver++: An Accurate Neural Solver for PDEs on Million-Scale Geometries}, 
      author={Huakun Luo and Haixu Wu and Hang Zhou and Lanxiang Xing and Yichen Di and Jianmin Wang and Mingsheng Long},
      year={2025},
      eprint={2502.02414},
      archivePrefix={arXiv},
      primaryClass={cs.LG}
}

@inproceedings{qi2017pointnet,
  author       = {Charles Ruizhongtai Qi and
                  Hao Su and
                  Kaichun Mo and
                  Leonidas J. Guibas},
  title        = {PointNet: Deep Learning on Point Sets for 3D Classification and Segmentation},
  booktitle    = {2017 {IEEE} Conference on Computer Vision and Pattern Recognition,
                  {CVPR} 2017, Honolulu, HI, USA, July 21-26, 2017},
  pages        = {77--85},
  publisher    = {{IEEE} Computer Society},
  year         = {2017},
  url          = {https://doi.org/10.1109/CVPR.2017.16},
  doi          = {10.1109/CVPR.2017.16},
  bibsource    = {dblp computer science bibliography, https://dblp.org}
}

@misc{ashton2025drivaerml,
      title={DrivAerML: High-Fidelity Computational Fluid Dynamics Dataset for Road-Car External Aerodynamics}, 
      author={Neil Ashton and Charles Mockett and Marian Fuchs and Louis Fliessbach and Hendrik Hetmann and Thilo Knacke and Norbert Schonwald and Vangelis Skaperdas and Grigoris Fotiadis and Astrid Walle and Burkhard Hupertz and Danielle Maddix},
      year={2025},
      eprint={2408.11969},
      archivePrefix={arXiv},
      primaryClass={physics.flu-dyn}
}

@misc{bonnet2023airfrans,
      author={Florent Bonnet and
              Ahmed Jocelyn Mazari and
              Paola Cinnella and 
              Patrick Gallinari},
      title={AirfRANS: High Fidelity Computational Fluid Dynamics Dataset for Approximating Reynolds-Averaged Navier-Stokes Solutions}, 
      year={2023},
      eprint={2212.07564},
      archivePrefix={arXiv},
      primaryClass={cs.LG}
}

@misc{alkin2025abuptscalingneuralcfd,
      title={AB-UPT: Scaling Neural CFD Surrogates for High-Fidelity Automotive Aerodynamics Simulations via Anchored-Branched Universal Physics Transformers}, 
      author={Benedikt Alkin and Maurits Bleeker and Richard Kurle and Tobias Kronlachner and Reinhard Sonnleitner and Matthias Dorfer and Johannes Brandstetter},
      year={2025},
      eprint={2502.09692},
      archivePrefix={arXiv},
      primaryClass={cs.LG}
}

@article{openfoam,
    author = {Weller, H. G. and Tabor, G. and Jasak, H. and Fureby, C.},
    title = {A tensorial approach to computational continuum mechanics using object-oriented techniques},
    journal = {Computer in Physics},
    volume = {12},
    number = {6},
    pages = {620-631},
    year = {1998},
    month = {11},
    issn = {0894-1866},
    doi = {10.1063/1.168744},
    eprint = {https://pubs.aip.org/aip/cip/article-pdf/12/6/620/7865493/620\_1\_online.pdf},
}

@InProceedings{simpleAlgo,
author="Caretto, L. S.
and Gosman, A. D.
and Patankar, S. V.
and Spalding, D. B.",
editor="Cabannes, Henri
and Temam, Roger",
title="Two calculation procedures for steady, three-dimensional flows with recirculation",
booktitle="Proceedings of the Third International Conference on Numerical Methods in Fluid Mechanics",
year="1973",
publisher="Springer Berlin Heidelberg",
address="Berlin, Heidelberg",
pages="60--68",
isbn="978-3-540-38392-5"
}

@book{kundu,
author = {Kundu, Pijush K. and Cohen, Ira M.},
address = {Amsterdam ;},
booktitle = {Fluid mechanics},
edition = {4th ed.},
isbn = {9780123737359},
language = {eng},
lccn = {2007042765},
publisher = {Academic Press},
title = {Fluid mechanics / Pijush K. Kundu, Ira M. Cohen ; with contributions by P.S. Ayyaswamy and H.H. Hu.},
year = {2008},
}

@book{Anderson2017,
  author = {Anderson, John D. Jr.},
  title = {Fundamentals of Aerodynamics},
  edition = {6th},
  publisher = {McGraw-Hill Education},
  year = {2017},
  address = {New York}
}

@article{brunton2020Review,
   author = "Brunton, Steven L. and Noack, Bernd R. and Koumoutsakos, Petros",
   title = "Machine Learning for Fluid Mechanics", 
   journal= "Annual Review of Fluid Mechanics",
   year = "2020",
   volume = "52",
   number = "Volume 52, 2020",
   pages = "477-508",
   doi = "https://doi.org/10.1146/annurev-fluid-010719-060214",
   publisher = "Annual Reviews",
   issn = "1545-4479",
   type = "Journal Article",
  }

@inproceedings{dosovitsky2021vit,
	author = {Alexey Dosovitskiy and Lucas Beyer and Alexander Kolesnikov and Dirk Weissenborn and Xiaohua Zhai and Thomas Unterthiner and Mostafa Dehghani and Matthias Minderer and Georg Heigold and Sylvain Gelly and Jakob Uszkoreit and Neil Houlsby},
	booktitle = {{ICLR}},
	title = {An Image is Worth 16x16 Words: Transformers for Image Recognition at Scale},
	year = {2021}}

@article{Peebles2022ScalableDM,
  title={Scalable Diffusion Models with Transformers},
  author={William S. Peebles and Saining Xie},
  journal={2023 IEEE/CVF International Conference on Computer Vision (ICCV)},
  year={2022},
  pages={4172-4182},
}

@article{sung_blendednet_2025,
  author       = {Nicholas Sung and
                  Steven Spreizer and
                  Mohamed Elrefaie and
                  Kaira M. Samuel and
                  Matthew C. Jones and
                  Faez Ahmed},
  title        = {BlendedNet: {A} Blended Wing Body Aircraft Dataset and Surrogate Model
                  for Aerodynamic Predictions},
  journal      = {CoRR},
  volume       = {abs/2509.07209},
  year         = {2025},
  doi          = {10.48550/ARXIV.2509.07209},
  eprinttype    = {arXiv},
  eprint       = {2509.07209},
  bibsource    = {dblp computer science bibliography, https://dblp.org}
}

@misc{kanchi_unifoil_2025,
      title={UniFoil: A Universal Dataset of Airfoils in Transitional and Turbulent Regimes for Subsonic and Transonic Flows}, 
      author={Rohit Sunil Kanchi and Benjamin Melanson and Nithin Somasekharan and Shaowu Pan and Sicheng He},
      year={2025},
      eprint={2505.21124},
      archivePrefix={arXiv},
      primaryClass={physics.flu-dyn},
}

@dataset{schillaci_airfoilcfd_2021,
  author       = {Schillaci, Alessandro and Quadrio, Maurizio and Boracchi, Giacomo},
  title        = {A database of CFD-computed flow fields around airfoils for machine-learning applications},
  year         = {2021},
  publisher    = {Zenodo},
  doi          = {10.5281/zenodo.4106752},
  note         = {Version 1.0}
}

@misc{ramos_airfoilcfd_2023,
title = {Airfoil Computational Fluid Dynamics - 9k shapes, 2 AoA's},
author = {Ramos, Dakota and Glaws, Andrew and King, Ryan and Lee, Bumseok and Doronina, Olga and Baeder, James and Vijayakumar, Ganesh and Grey, Zachary},
year = {2023},
howpublished = {Open Energy Data Initiative (OEDI), National Renewable Energy Laboratory (NREL), https://doi.org/10.25984/2222587},
note = {Accessed: 2025-10-30},
doi = {10.25984/2222587}
}

@article{alkin_luminary_2025,
  title={AB-UPT for Automotive and Aerospace Applications},
  author={Alkin, Benedikt and Kurle, Richard and Serrano, Louis and Just, Dennis and Brandstetter, Johannes},
  journal={arXiv preprint arXiv:2510.15808},
  year={2025}
}

@inproceedings{chen_lion_2023,
  author       = {Xiangning Chen and
                  Chen Liang and
                  Da Huang and
                  Esteban Real and
                  Kaiyuan Wang and
                  Hieu Pham and
                  Xuanyi Dong and
                  Thang Luong and
                  Cho{-}Jui Hsieh and
                  Yifeng Lu and
                  Quoc V. Le},
  editor       = {Alice Oh and
                  Tristan Naumann and
                  Amir Globerson and
                  Kate Saenko and
                  Moritz Hardt and
                  Sergey Levine},
  title        = {Symbolic Discovery of Optimization Algorithms},
  booktitle    = {Advances in Neural Information Processing Systems 36: Annual Conference
                  on Neural Information Processing Systems 2023, NeurIPS 2023, New Orleans,
                  LA, USA, December 10 - 16, 2023},
  year         = {2023},
  bibsource    = {dblp computer science bibliography, https://dblp.org}
}

@inproceedings{devlin_bert_2019,
  author       = {Jacob Devlin and
                  Ming{-}Wei Chang and
                  Kenton Lee and
                  Kristina Toutanova},
  editor       = {Jill Burstein and
                  Christy Doran and
                  Thamar Solorio},
  title        = {{BERT:} Pre-training of Deep Bidirectional Transformers for Language
                  Understanding},
  booktitle    = {Proceedings of the 2019 Conference of the North American Chapter of
                  the Association for Computational Linguistics: Human Language Technologies,
                  {NAACL-HLT} 2019, Minneapolis, MN, USA, June 2-7, 2019, Volume 1 (Long
                  and Short Papers)},
  pages        = {4171--4186},
  publisher    = {Association for Computational Linguistics},
  year         = {2019},
  doi          = {10.18653/V1/N19-1423},
  bibsource    = {dblp computer science bibliography, https://dblp.org}
}

@misc{ Alletto2024,
   author = "OpenFOAM Wiki",
   title = "OneraM6 by Michael Alletto --- OpenFOAM Wiki{,} ",
   year = "2024",
   url = "https://wiki.openfoam.com/index.php?title=OneraM6_by_Michael_Alletto&oldid=4394",
   note = "[Online; accessed 12-November-2025]"
 }

@article{alkin2024universal,
  title={Universal physics transformers: A framework for efficiently scaling neural operators},
  author={Alkin, Benedikt and F{\"u}rst, Andreas and Schmid, Simon and Gruber, Lukas and Holzleitner, Markus and Brandstetter, Johannes},
  journal={Advances in Neural Information Processing Systems},
  volume={37},
  pages={25152--25194},
  year={2024}
}

@article{
li2023transformer,
title={Transformer for Partial Differential Equations{\textquoteright} Operator Learning},
author={Zijie Li and Kazem Meidani and Amir Barati Farimani},
journal={Transactions on Machine Learning Research},
issn={2835-8856},
year={2023},
url={https://openreview.net/forum?id=EPPqt3uERT},
note={}
}

@inproceedings{
pfaff2021learning,
title={Learning Mesh-Based Simulation with Graph Networks},
author={Tobias Pfaff and Meire Fortunato and Alvaro Sanchez-Gonzalez and Peter Battaglia},
booktitle={International Conference on Learning Representations},
year={2021}
}

@article{li2020neural,
  title={Neural operator: Graph kernel network for partial differential equations},
  author={Li, Zongyi and Kovachki, Nikola and Azizzadenesheli, Kamyar and Liu, Burigede and Bhattacharya, Kaushik and Stuart, Andrew and Anandkumar, Anima},
  journal={arXiv preprint arXiv:2003.03485},
  year={2020}
}

@inproceedings{teney2024neural,
  title={Neural redshift: Random networks are not random functions},
  author={Teney, Damien and Nicolicioiu, Armand Mihai and Hartmann, Valentin and Abbasnejad, Ehsan},
  booktitle={Proceedings of the IEEE/CVF Conference on Computer Vision and Pattern Recognition},
  pages={4786--4796},
  year={2024}
}

@article{koehler2025neural,
  title={Neural Emulator Superiority: When Machine Learning for PDEs Surpasses its Training Data},
  author={Koehler, Felix and Thuerey, Nils},
  journal={arXiv preprint arXiv:2510.23111},
  year={2025}
}

@article{Hansen2001CMAES,
author = {Hansen, Nikolaus and Ostermeier, Andreas},
title = {Completely Derandomized Self-Adaptation in Evolution Strategies},
year = {2001},
issue_date = {June 2001},
publisher = {MIT Press},
address = {Cambridge, MA, USA},
volume = {9},
number = {2},
issn = {1063-6560},
url = {https://doi.org/10.1162/106365601750190398},
doi = {10.1162/106365601750190398},
journal = {Evol. Comput.},
month = jun,
pages = {159–195},
numpages = {37}
}

@misc{Hansen2023cmaes,
      title={The CMA Evolution Strategy: A Tutorial}, 
      author={Nikolaus Hansen},
      year={2023},
      eprint={1604.00772},
      archivePrefix={arXiv},
      primaryClass={cs.LG},
      url={https://arxiv.org/abs/1604.00772}, 
}

@inproceedings{kingma2014adam,
  title={Adam: A Method for Stochastic Optimization},
  author={Kingma, Diederik P. and Ba, Jimmy},
  booktitle={International Conference on Learning Representations (ICLR)},
  year={2015}
}

@article{Jones1998EfficientGO,
  title={Efficient Global Optimization of Expensive Black-Box Functions},
  author={Donald R. Jones and Matthias Schonlau and William J. Welch},
  journal={Journal of Global Optimization},
  year={1998},
  volume={13},
  pages={455-492},
  url={https://api.semanticscholar.org/CorpusID:263864014}
}

@inproceedings{Schmitt1979PressureDO,
  title={Pressure distributions on the ONERA M6 wing at transonic Mach numbers},
  author={V. Schmitt and Françoise Charpin},
  year={1979}
}

@inbook{mani_oneram6_1997,
author = {M. Mani and J. Ladd and A. Cain and R. Bush and M. Mani and J. Ladd and A. Cain and R. Bush},
title = {An assessment of one- and two-equation turbulence models for internal and external flows},
booktitle = {28th Fluid Dynamics Conference},
doi = {10.2514/6.1997-2010},
eprint = {https://arc.aiaa.org/doi/pdf/10.2514/6.1997-2010},
year={1997}
}

@article{li_data-based_2021,
title = {Data-based approach for wing shape design optimization},
journal = {Aerospace Science and Technology},
volume = {112},
pages = {106639},
year = {2021},
issn = {1270-9638},
doi = {https://doi.org/10.1016/j.ast.2021.106639},
url = {https://www.sciencedirect.com/science/article/pii/S1270963821001498},
author = {Jichao Li and Mengqi Zhang}
}

@inproceedings{martins2020perspectives,
  title={Perspectives on aerodynamic design optimization},
  author={Martins, Joaquim RRA},
  booktitle={AIAA Scitech 2020 Forum},
  pages={0043},
  year={2020}
}

@article{lyu_aerodynamic_2015,
author = {Lyu, Zhoujie and Kenway, Gaetan K. W. and Martins, Joaquim R. R. A.},
title = {Aerodynamic Shape Optimization Investigations of the Common Research Model Wing Benchmark},
journal = {AIAA Journal},
volume = {53},
number = {4},
pages = {968-985},
year = {2015},
doi = {10.2514/1.J053318},
}

@article{he_robust_2019,
title = {Robust aerodynamic shape optimization—From a circle to an airfoil},
journal = {Aerospace Science and Technology},
volume = {87},
pages = {48-61},
year = {2019},
issn = {1270-9638},
doi = {https://doi.org/10.1016/j.ast.2019.01.051},
author = {Xiaolong He and Jichao Li and Charles A. Mader and Anil Yildirim and Joaquim R.R.A. Martins}
}

@article{li_efficient_2020,
author = {Li, Jichao and Zhang, Mengqi and Martins, Joaquim R. R. A. and Shu, Chang},
title = {Efficient Aerodynamic Shape Optimization with Deep-Learning-Based Geometric Filtering},
journal = {AIAA Journal},
volume = {58},
number = {10},
pages = {4243-4259},
year = {2020},
doi = {10.2514/1.J059254}
}

@article{bartoli_adaptive_2019,
title = {Adaptive modeling strategy for constrained global optimization with application to aerodynamic wing design},
journal = {Aerospace Science and Technology},
volume = {90},
pages = {85-102},
year = {2019},
issn = {1270-9638},
doi = {https://doi.org/10.1016/j.ast.2019.03.041},
url = {https://www.sciencedirect.com/science/article/pii/S1270963818306011},
author = {N. Bartoli and T. Lefebvre and S. Dubreuil and R. Olivanti and R. Priem and N. Bons and J.R.R.A. Martins and J. Morlier}
}

@article{benaouali_multidisciplinary_2019,
title = {Multidisciplinary design optimization of aircraft wing using commercial software integration},
journal = {Aerospace Science and Technology},
volume = {92},
pages = {766-776},
year = {2019},
issn = {1270-9638},
doi = {https://doi.org/10.1016/j.ast.2019.06.040},
author = {Abdelkader Benaouali and Stanisław Kachel}
}

@article{shi_natural_2020,
author = {Shi, Yayun and Mader, Charles A. and He, Sicheng and Halila, Gustavo L. O. and Martins, Joaquim R. R. A.},
title = {Natural Laminar-Flow Airfoil Optimization Design Using a Discrete Adjoint Approach},
journal = {AIAA Journal},
volume = {58},
number = {11},
pages = {4702-4722},
year = {2020},
doi = {10.2514/1.J058944}
}

@article{kenway_effective_2019,
title = {Effective adjoint approaches for computational fluid dynamics},
journal = {Progress in Aerospace Sciences},
volume = {110},
pages = {100542},
year = {2019},
issn = {0376-0421},
doi = {https://doi.org/10.1016/j.paerosci.2019.05.002},
url = {https://www.sciencedirect.com/science/article/pii/S0376042119300120},
author = {Gaetan K.W. Kenway and Charles A. Mader and Ping He and Joaquim R.R.A. Martins}
}

\newpage
\appendix

\section{Drag/Lift coefficients}
\label{app:drag_lift_formulation}

In this section we provide a definition of the drag and lift coefficients, commonly used in aerospace engineering.
Aerodynamic forces quantify interesting properties of a wing such as drag and lift, which are the result of an integration of the surface pressure $p_s$ and friction $\bm{\tau}$, i.e., force per unit area exerted by the fluid on the surface, acting tangential (parallel) to the surface.
These forces are obtained from the total force acting on an object in an airflow, which is given by
\begin{align}
    \bm{F} = \oint_S \bigg( -(p_s-p_\infty) \bm{n} + \bm{\tau} \bigg) dS \ ,
\end{align}
where $p_s$ is the surface pressure, $p_\infty$ the free stream pressure, $\bm{n}$ the surface normal vector, and $\bm{\tau}$ the contribution of friction.
Drag and lift coefficients are defined as dimensionless numbers
\begin{equation}
    C_\text{D} =   \frac{2\, \bm{F} \cdot \bm{e}_{\text{drag}}}{\rho\, v^2 A_\text{ref}}, \; C_\text{l} =   \frac{2\, \bm{F} \cdot \bm{e}_\text{lift}}{\rho\, v^2 A_\text{ref}},
    \label{eq:drag_and_lift_coefficient}
\end{equation}
and often used in engineering~\citep{ashton2025drivaerml}, where $\bm{e}_{\text{drag}}$ is a unit vector into the free stream direction, $\bm{e}_{\text{lift}}$ a unit vector into the lift direction perpendicular to the free stream direction, $\rho$ the density, $v$ the magnitude of the free stream velocity, and $A_{\text{ref}}$ a characteristic reference area for a certain geometry.
Further, the acting force $\bm{F}$ can be converted into dimensionless numbers as drag and lift forces

\begin{equation}
    \bm{F}_\text{drag} =   \bm{F} \cdot \bm{e}_{\text{drag}}, \; \bm{F}_\text{lift} =   \bm{F} \cdot \bm{e}_\text{lift} \ .
    \label{eq:drag_and_lift_force}
\end{equation}

When simulating an airplane in flight, the angle of attack $\alpha$ is a common boundary condition which changes free stream $\bm{e}_\text{drag}$ and lift direction $\bm{e}_\text{lift}$ as given by the unit velocity vector $\bm{u}_\infty$ as
\begin{equation}
    \bm{u}_\infty = \begin{bmatrix} \cos(\alpha) \\ 0 \\ \sin(\alpha) \end{bmatrix}, \;
    \bm{e}_\text{drag} = \frac{\bm{u}_\infty}{\|\bm{u}_\infty\|_2}  , 
    \; \bm{e}_\text{lift} = \bm{e}_\text{drag} \times \begin{bmatrix} 0 \\ 1 \\ 0 \end{bmatrix} \ , 
    \label{eq:drag_and_lift_direction}
\end{equation}

where $\|.\|_2$ is the Euclidian norm. For $\alpha=0$ these reduce to $\bm{e}_\text{drag} = (1, 0, 0)$ and $\bm{e}_\text{lift} = (0, 0, 1)$.

\section{Implementation Details}
\label{app:implementation_details}

We perform a hyperparameter search over learning rate $\texttt{lr} \in \{ 1e-5, 3e-5, 5e-5, 1e-4 \}$ for all methods and select the best performing one.
We train for 10 epochs with a learning rate of $1e-5$ with linear warmup for the first $5\%$ of training and a cosine decay thereafter. 
We train in \texttt{float16} precision using the Lion optimizer \cite{chen_lion_2023} with a weight decay of $\lambda=0.05$ and gradient clipping of 0.25. 
We preprocess the different fields using z-score normalization except for vorticity where we apply normalization by the average magnitude.
We elaborate on the method-specific details for the different methods as follows.

\textbf{AB-UPT.}
For AB-UPT we use a hidden dimension of 192 and and set the number of anchor points for both volume and surface branches to $16,384$.
Before processing the geometry we subsample it to $65,536$ points and perform supernode pooling to $16,384$ points.
The hidden dimensionality is set to 192 and the total number of parameters for this model amounts to 7.1 million.

\textbf{Transformer.}
We use a Transformer architecture based on coordinates encoded via continuous sine-cosine positional embeddings \citep{devlin_bert_2019} and a hidden dimension of 192.
We employ 16 Transformer blocks with three attention heads, resulting in a parameter count of roughly 7.4M parameters.

\textbf{Transolver.}
The Transolver baseline integrates the attention mechanism from \citep{wu2024Transolver} into the Transformer baseline. 
We use 512 slices for the Transolver attention.
The remainder of the architecture is the same as for the Transformer baseline, resulting in a parameter count of 7.5M parameters.

\textbf{PointNet.}
We use a hidden dimension of 192 and an input embedding of 96. We use a global position embedding of 3072 dimensions, resulting in a model size of roughly 8.3M parameters.

\section{Ablation studies}
\label{app:ablations}

Our dataset incorporates variations in geometric design parameters that are \emph{observable} through the point cloud and \emph{non-observable} inflow conditions (e.g., inflow Mach number). 
We therefore performed an ablation study on the best-performing model, AB-UPT, to assess the impact of conditioning on these parameters (relative L2 error, Table \ref{tab:geometry_inflow_ablation}). 
The results demonstrate that the AB-UPT architecture effectively infers information about the geometry directly from the point cloud, while non-observable parameters like inflow conditions are required as explicit conditioning inputs.
\begin{table}[h!]
\centering
\caption{Impact of geometry and inflow conditioning on the AB-UPT surrogate model's performance, measured by relative L2 error (\%) on the \emph{in-distribution} test set. All models were trained for 10 epochs.}
\label{tab:geometry_inflow_ablation}
\begin{tabular}{llccccc}
\toprule
\multicolumn{2}{c}{Conditioning} & \multicolumn{5}{c}{\textbf{Relative L2 error} (\%)} \\
\cmidrule(lr){1-2} \cmidrule(lr){3-7}
Geometry & Inflow & $p_s$ & $\boldsymbol{\tau}$ & $p_v$ & $\boldsymbol{u}$ & $\boldsymbol{\omega}$ \\
\midrule
\ccmark & \ccmark & 0.53 & 4.16 & 0.49 & 3.36 & 10.5 \\
\ccmark & \cxmark & 0.54 & 4.16 & 0.51 & 3.42 & 10.6 \\
\cxmark & \ccmark & 10.8 & 43.5 & 9.58 & 29.5 & 33.7 \\
\cxmark & \cxmark & 10.9 & 43.7 & 9.59 & 29.6 & 34.3 \\
\bottomrule
\end{tabular}
\end{table}

\section{Additional results}
\label{app:results}

\begin{figure}
    \centering
    \begin{subfigure}[t]{0.49\textwidth}
        \includegraphics[width=\textwidth]{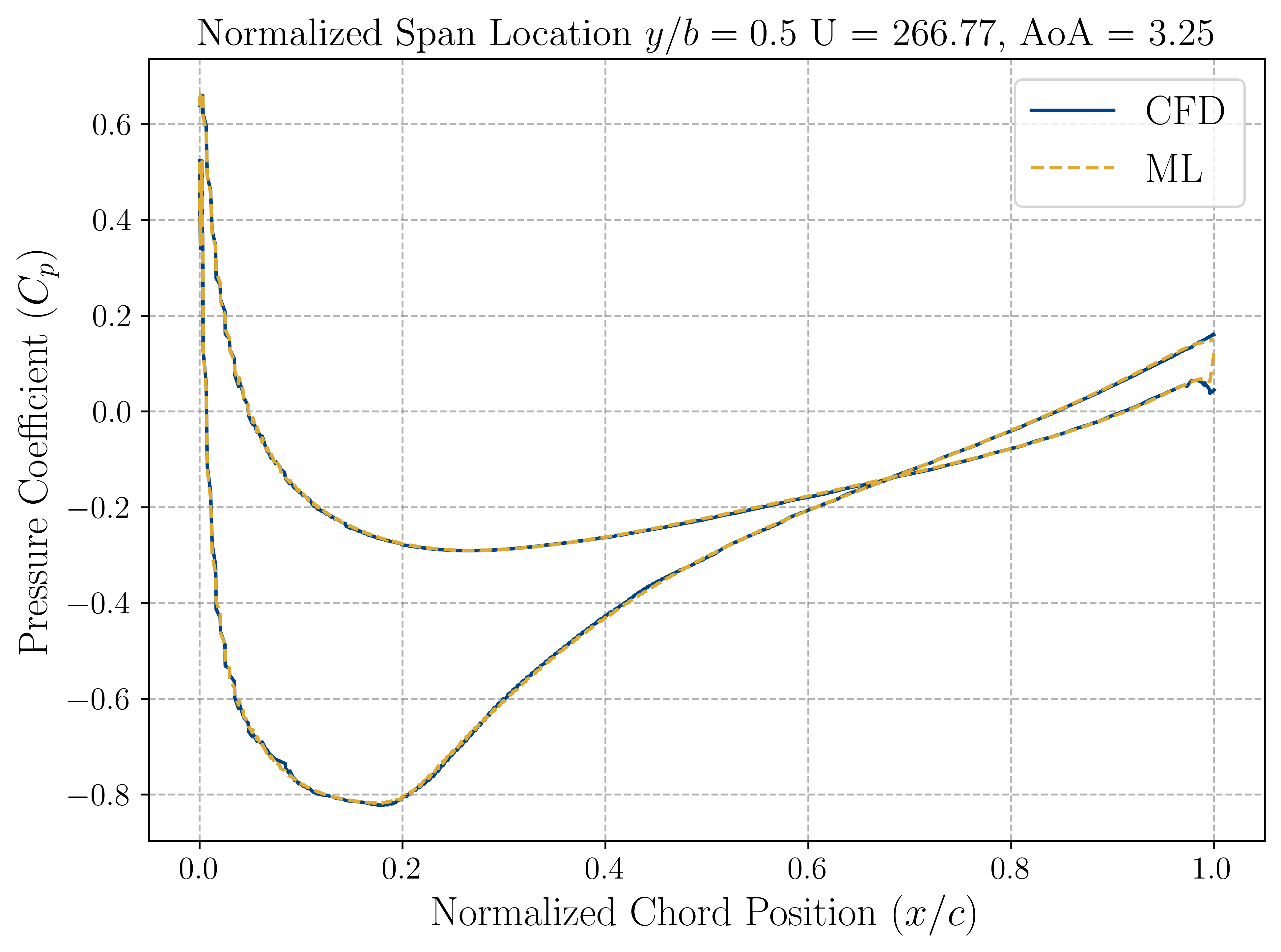}
    \end{subfigure}
    \hfill
    \begin{subfigure}[t]{0.49\textwidth}
        \includegraphics[width=\textwidth]{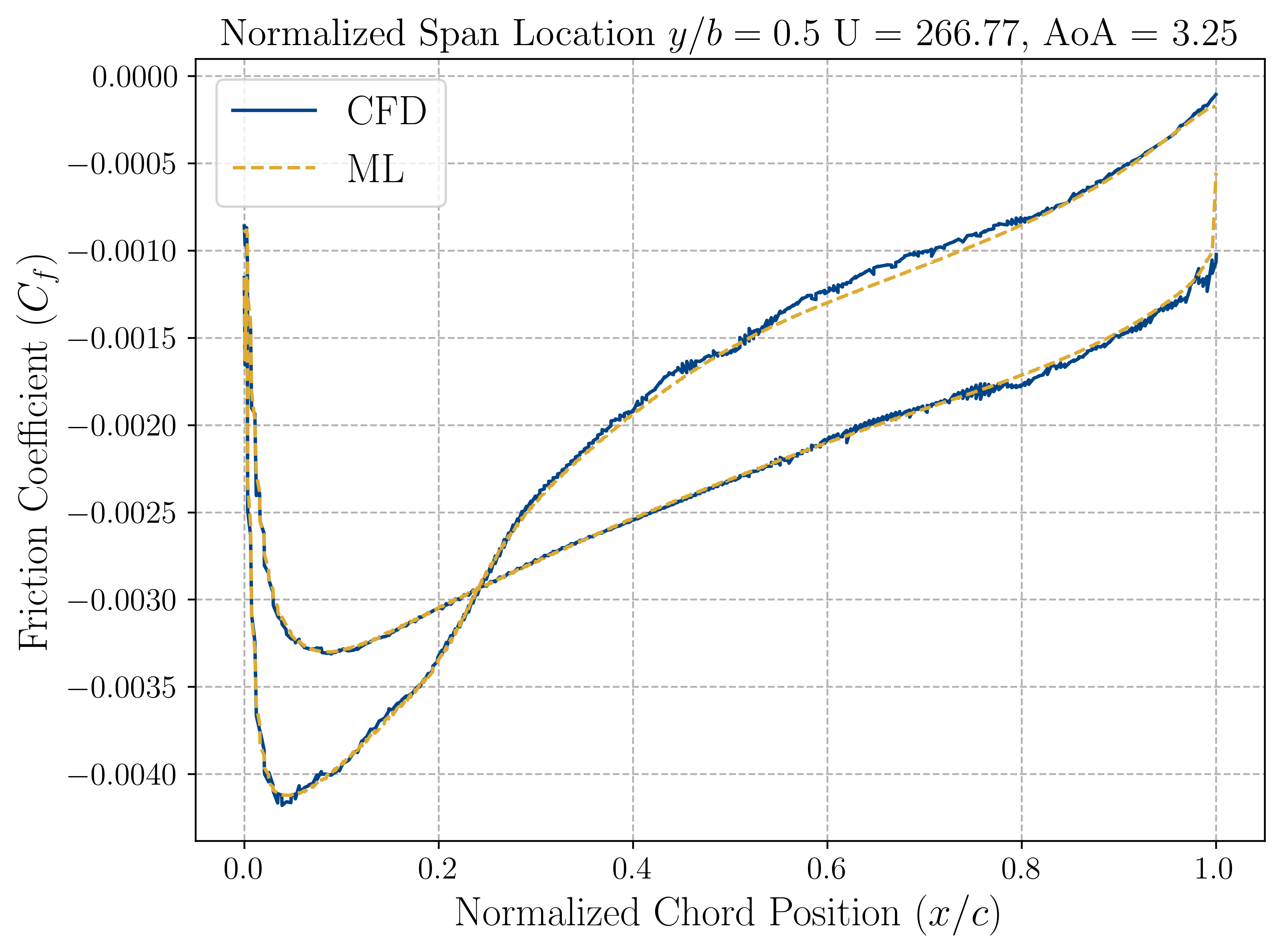}
    \end{subfigure}
    \caption{Comparison between CFD data (blue) and AB-UPT prediction (dashed yellow) for the pressure coefficient $(C_p)$ and friction coefficient $(C_f)$ on the wing surface at a normalized span location of $y/b=0.5$ for a case of the OOD test set. 3D visualizations are shown in Figure~\ref{fig:3D_comparison_test_extrapol_combined}.}
    \label{fig:profile_plot_test_extrapol}
\end{figure}

\begin{figure}
    \centering
    \begin{subfigure}[t]{0.49\textwidth}
        \includegraphics[width=\textwidth]{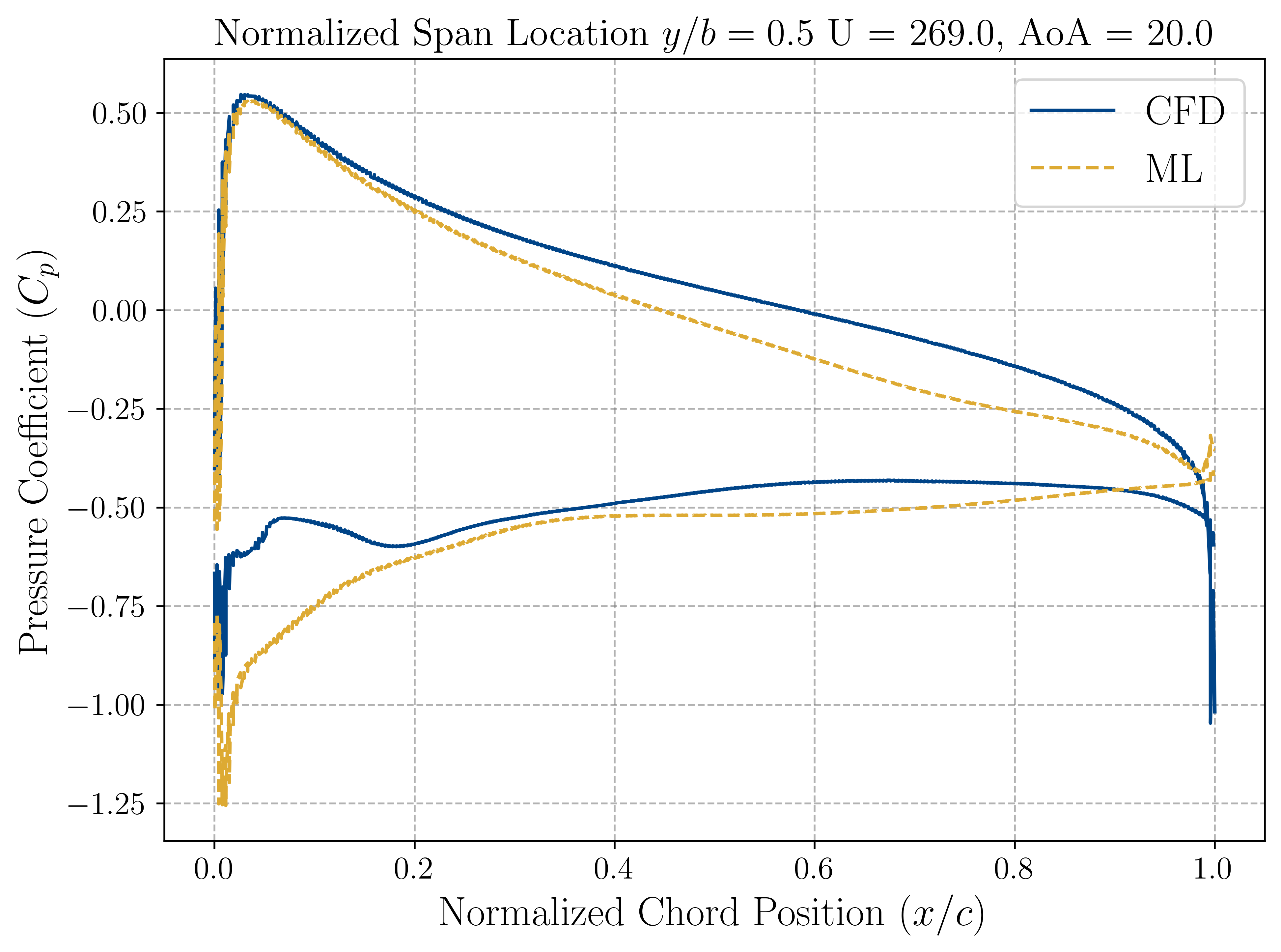}
        \label{fig:subfig1}
    \end{subfigure}
    \hfill
    \begin{subfigure}[t]{0.49\textwidth}
        \includegraphics[width=\textwidth]{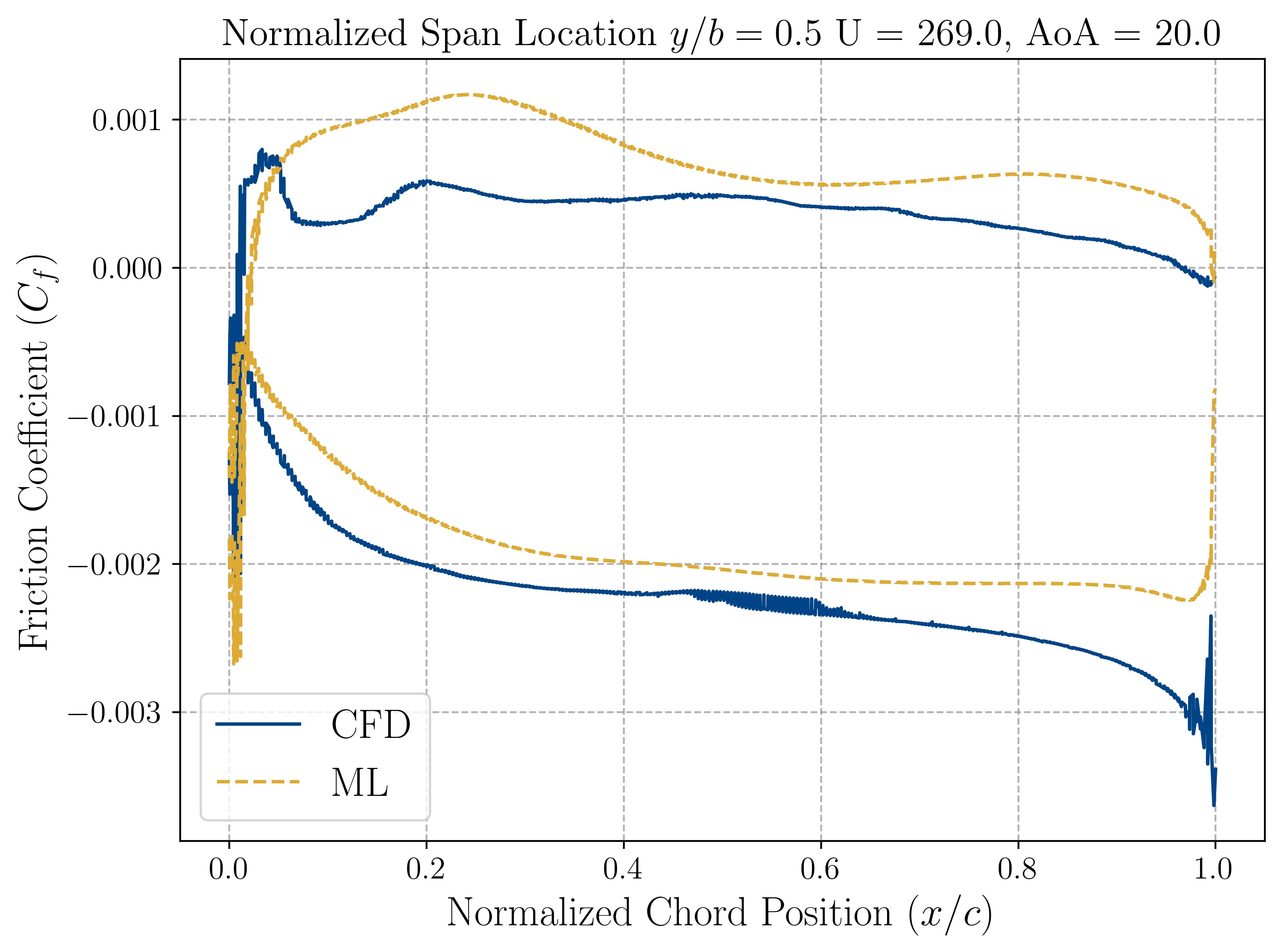}
    \end{subfigure}
    \caption{Comparison between CFD data (blue) and AB-UPT prediction (dashed yellow) for the pressure coefficient $(C_p)$ and friction coefficient $(C_f)$ on the wing surface at a normalized span location of $y/b=0.5$ for a case of the parameter scans. 3D visualizations are shown Figure~\ref{fig:3D_comparison_test_scans_combined}.}
    \label{fig:profile_plot_test_scans}
\end{figure}

\begin{figure}[t!]
    \centering
    \begin{subfigure}{1.0\linewidth}
        \centering
        \includegraphics[width=\linewidth]{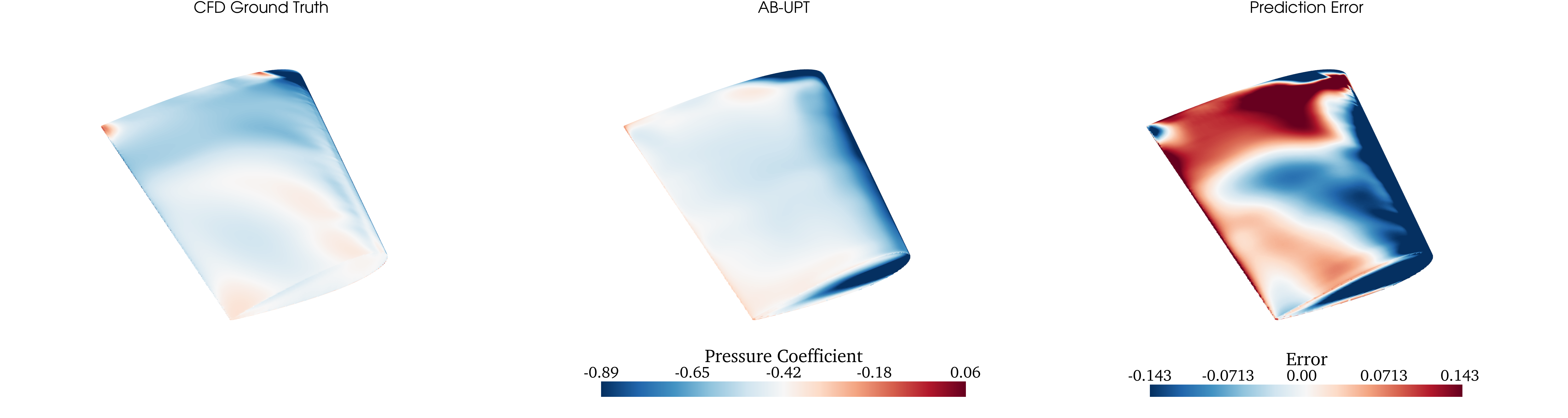}
        \caption{Pressure coefficient $(C_p)$}
        \label{fig:3D_comparison_test_scans_cp}
    \end{subfigure}
    \vspace{0.5cm}
    \begin{subfigure}{1.0\linewidth}
        \centering
        \includegraphics[width=\linewidth]{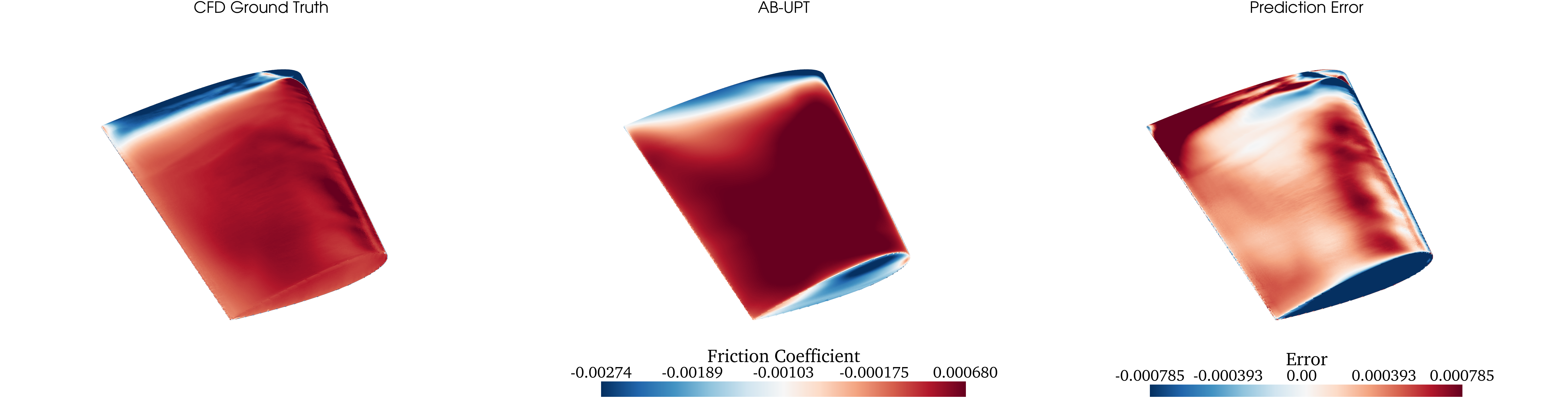}
        \caption{Friction coefficient $(C_f)$}
        \label{fig:3D_comparison_test_scans_cf}
    \end{subfigure}
    \caption{Comparison between surface field coefficients on the wing's surface of the CFD (left), AB-UPT surrogate (center) and the error between them (right). The case presented is from the the parameter scan with $\Lambda=40$ and $\alpha=20$ far outside the training range. Corresponding surface pressure and friction profile plots are shown in Figure~\ref{fig:profile_plot_test_scans}.}
    \label{fig:3D_comparison_test_scans_combined}
\end{figure}

In Figure \ref{fig:profile_plot_test_extrapol}, we show the pressure and friction coefficients predicted by AB-UPT for a normalized span location of $y/b=0.5$ and compare it to the CFD simulation, showing that AB-UPT yields highly accurate predictions.
We provide the same visualization for a different case with an angle-of-attack that is out of the training range ($\alpha=20^\circ$) in Figure \ref{fig:profile_plot_test_scans}.
Finally, we also show the 3D visualization of the corresponding wing including prediction error of the model in Figure \ref{fig:3D_comparison_test_scans_combined}.

\begin{figure}[t!]
    \centering
    \begin{subfigure}{0.32\linewidth}
        \centering
        \includegraphics[width=\linewidth]{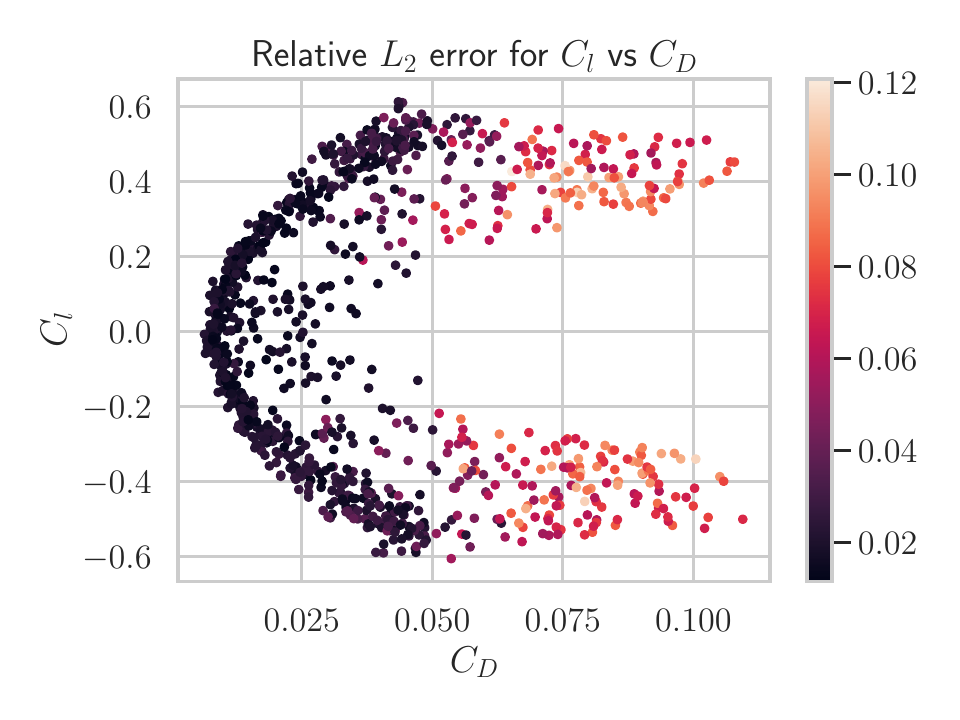}
    \end{subfigure}
    \hfill
    \begin{subfigure}{0.32\linewidth}
        \centering
        \includegraphics[width=\linewidth]{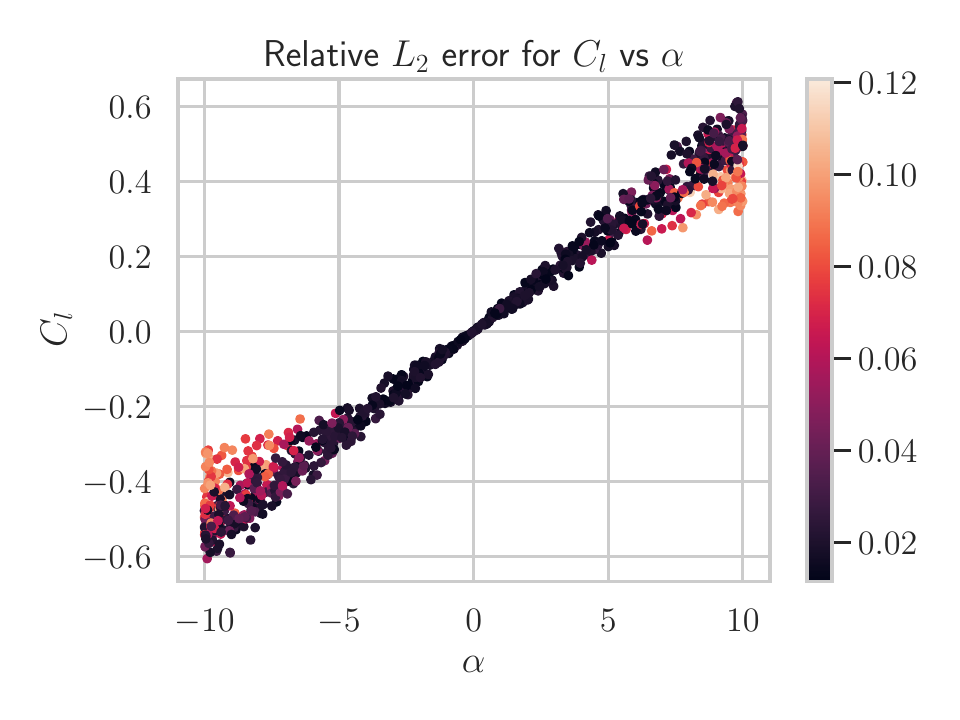}
    \end{subfigure}
    \hfill
    \begin{subfigure}{0.32\linewidth}
        \centering
        \includegraphics[width=\linewidth]{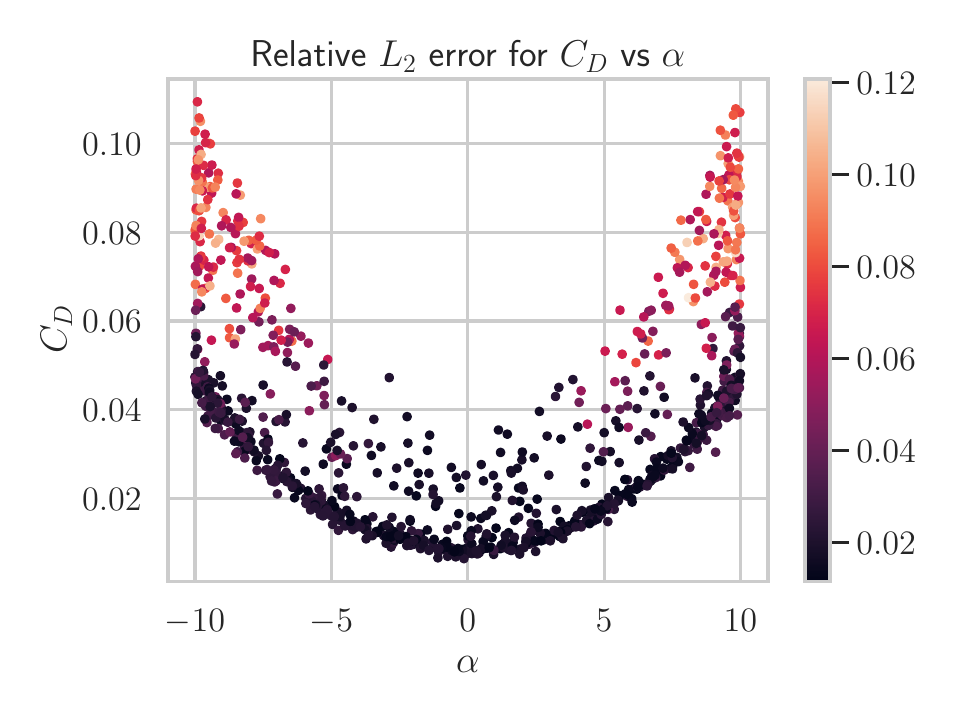}
    \end{subfigure}
    \caption{Distribution of relative $L_2$ error for $C_D$ versus  $C_l$ (left), $C_D$ versus $\alpha$ (middle), and $C_l$ versus $\alpha$ (right) for the OOD test set. Most error is introduced in high $C_D$ regimes, absent from the Pareto front.}
    \label{fig:average_l2err_test_extrapol}
\end{figure}

\begin{figure}[t!]
    \centering

    \begin{subfigure}{0.32\linewidth}
        \centering
        \includegraphics[width=\linewidth]{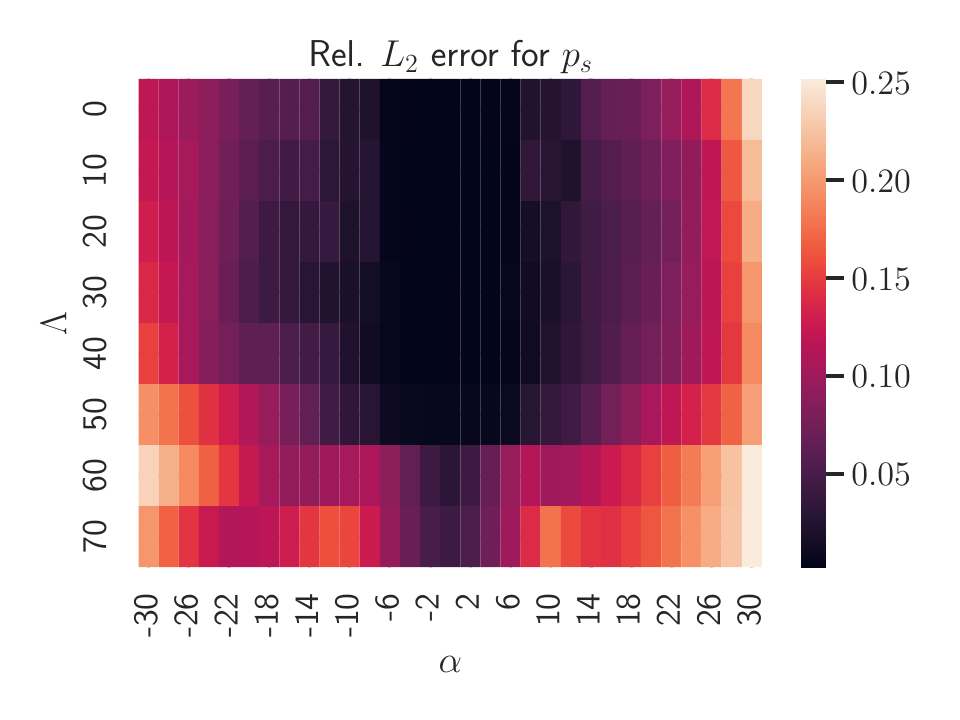}
    \end{subfigure}
    \hfill
    \begin{subfigure}{0.32\linewidth}
        \centering
        \includegraphics[width=\linewidth]{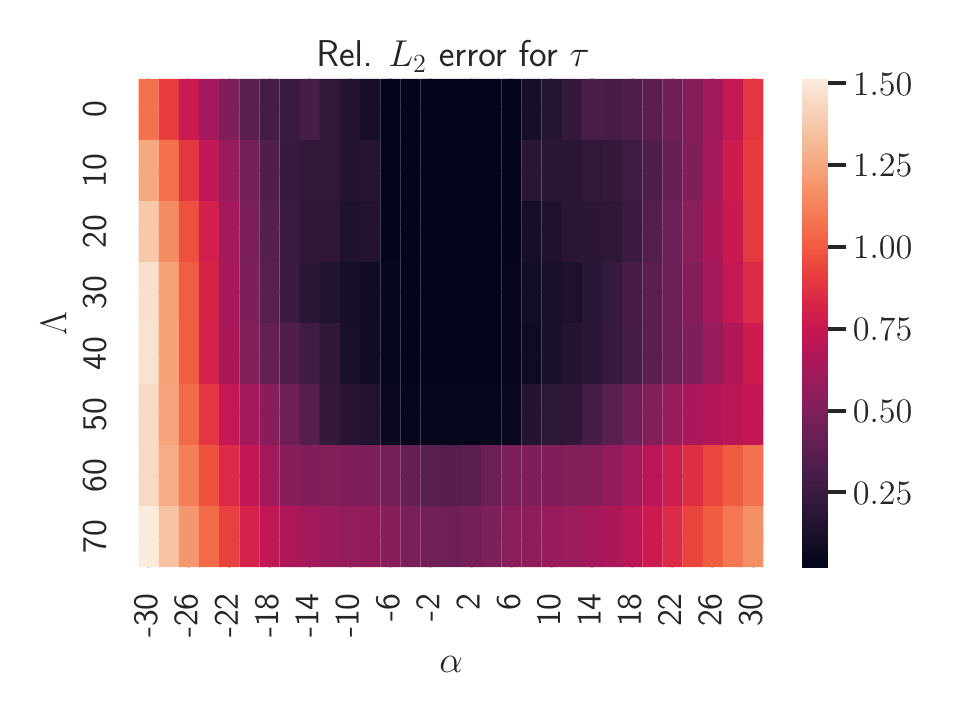}
    \end{subfigure}
    \hfill
    \begin{subfigure}{0.32\linewidth}
        \centering
        \includegraphics[width=\linewidth]{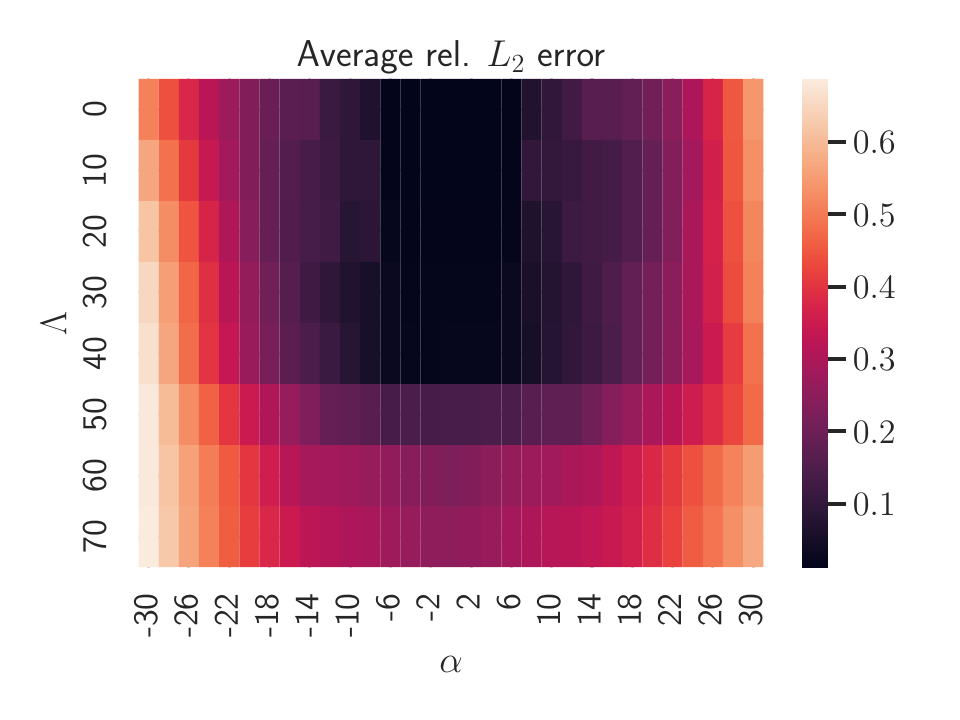}
    \end{subfigure}
    \caption{Relative $L_2$ error for $p_s$ (left), $\tau$ (middle), and averaged over both (right) for parameter scans over $\alpha$ and $\Lambda$. AB-UPT is most error-prone at out-of-range values for $\Lambda \in \{ 60,70\}$ and $\alpha \in \{ -26,-28,-30 \}$. For other out-of-range parameters AB-UPT is surprisingly stable.}
    \label{fig:errs_sweep_vs_aoa}
\end{figure}

\section{Data visualizations}
In this section, we visualize several cases obtained by our numerical simulations. An important aspect when generating a dataset is quality control to ensure that all simulations have converged to meaningful solutions. Establishing convergence criteria for transonic wing simulations is challenging. We employed classical metrics including  monitoring of residuals and force/moment coefficients on the wing. However, these indicators may not always distinguish between physically unsteady and/or discontinuous flows and numerical artifacts.

\begin{figure}
    \centering
    \begin{subfigure}[b]{0.49\textwidth}
        \includegraphics[width=\textwidth]{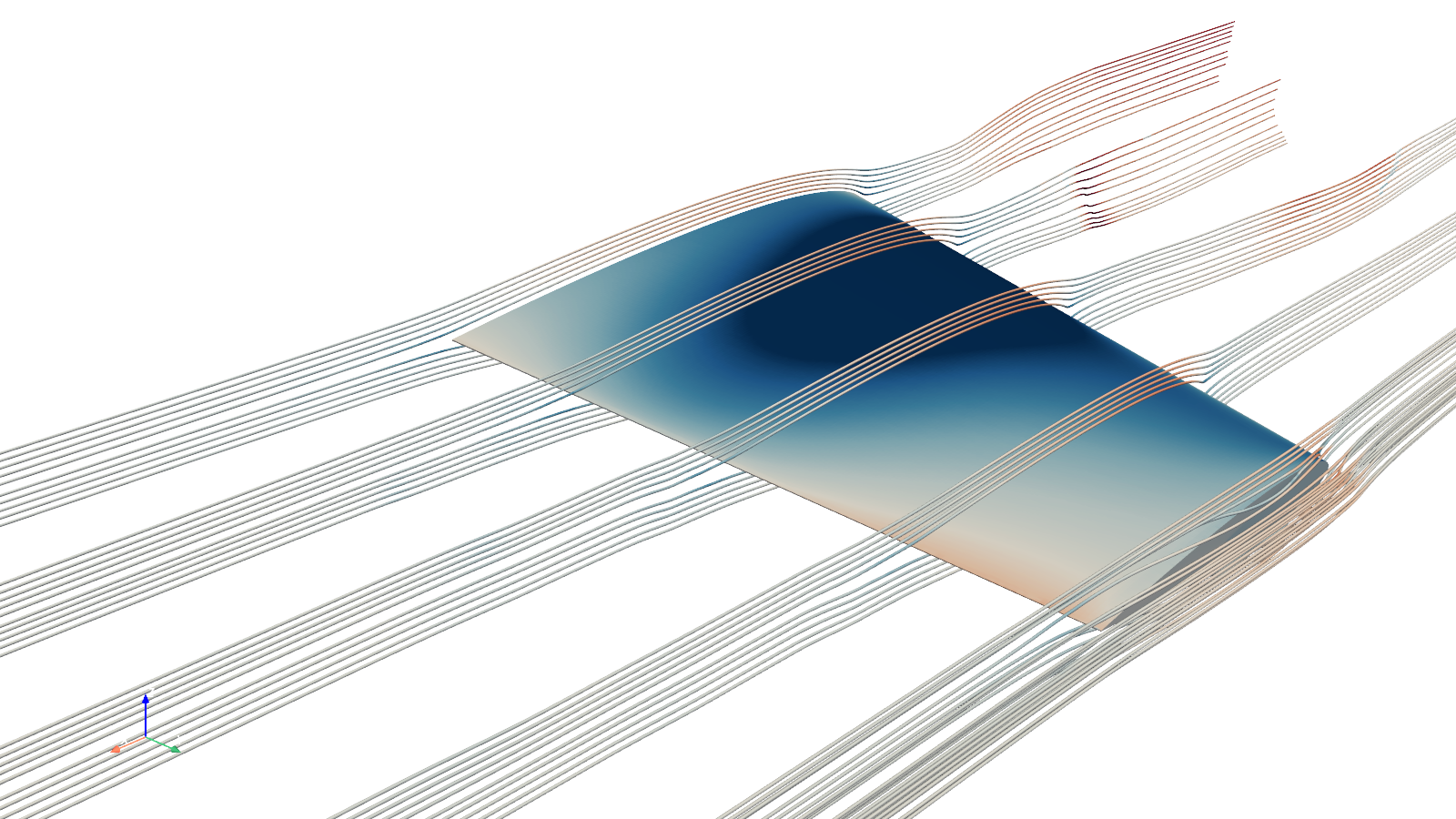}
        \label{fig:failed_case_1}
    \end{subfigure}
    \hfill
    \begin{subfigure}[b]{0.49\textwidth}
        \includegraphics[width=\textwidth]{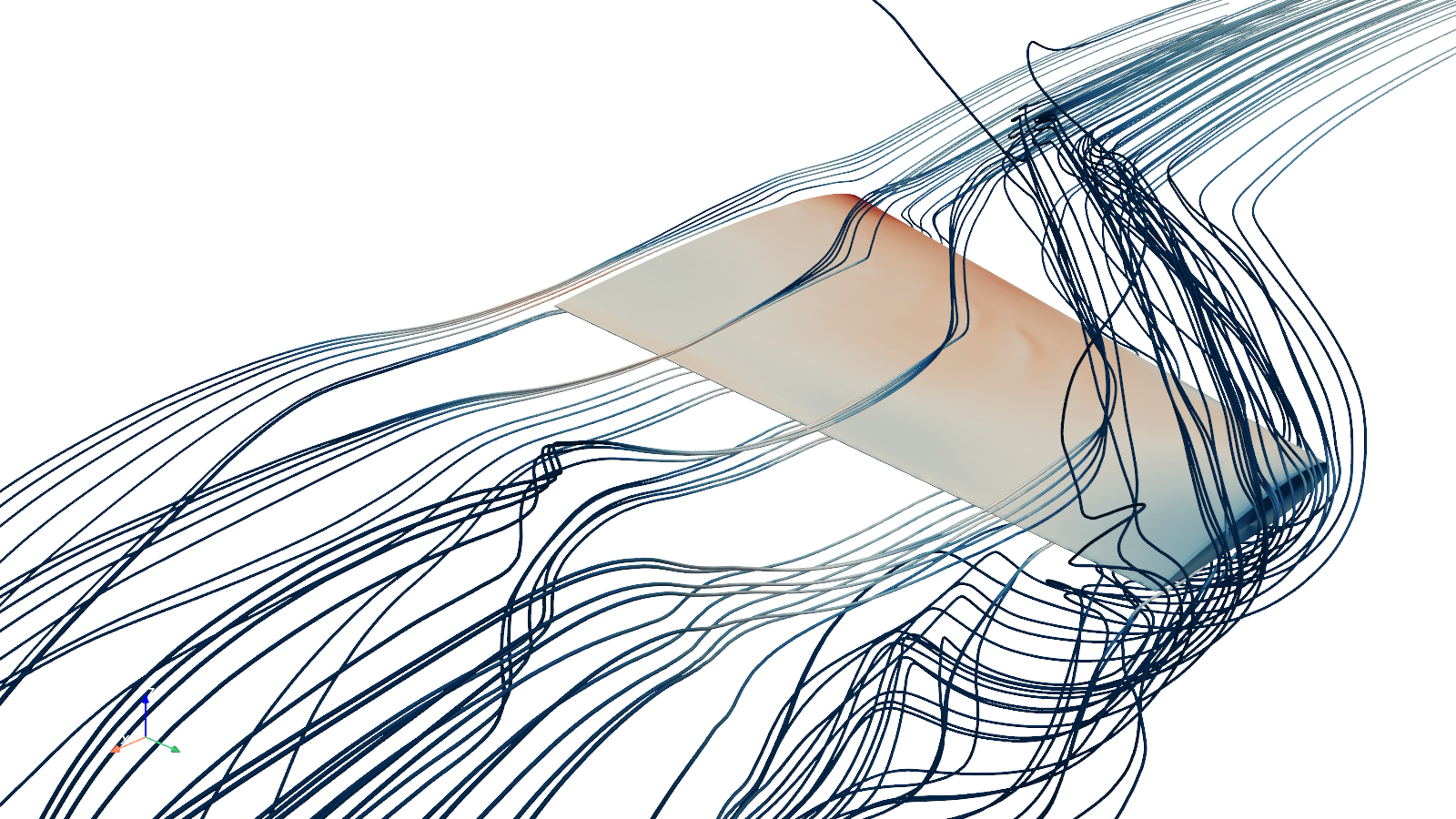}
        \label{fig:failed_case_2}
    \end{subfigure}
    \caption{Illustration of  failed cases caused by spanwise mesh quality variations (left) and numerically diverging case (right).}
    \label{fig:failed_cases}
\end{figure}

To supplement these conventional approaches, we employed AB-UPT as an additional quality control tool. The model's prediction error served as an indicator of data quality—cases with anomalously high errors were flagged as potentially unconverged. This approach successfully identified failed cases where classical convergence monitors had been ambiguous. Manual inspection confirmed these cases exhibited spurious flow patterns, inadequate boundary layer mesh resolution, or early numerical divergence (Figure \ref{fig:failed_cases}). Removing these contaminated samples ensured the final dataset comprised numerically consistent solutions.

\begin{figure}[!htbp]
    \centering
    \begin{subfigure}[b]{0.475\textwidth}
        \centering
        \includegraphics[width=\linewidth, trim={0 100pt 0 80pt}, clip]{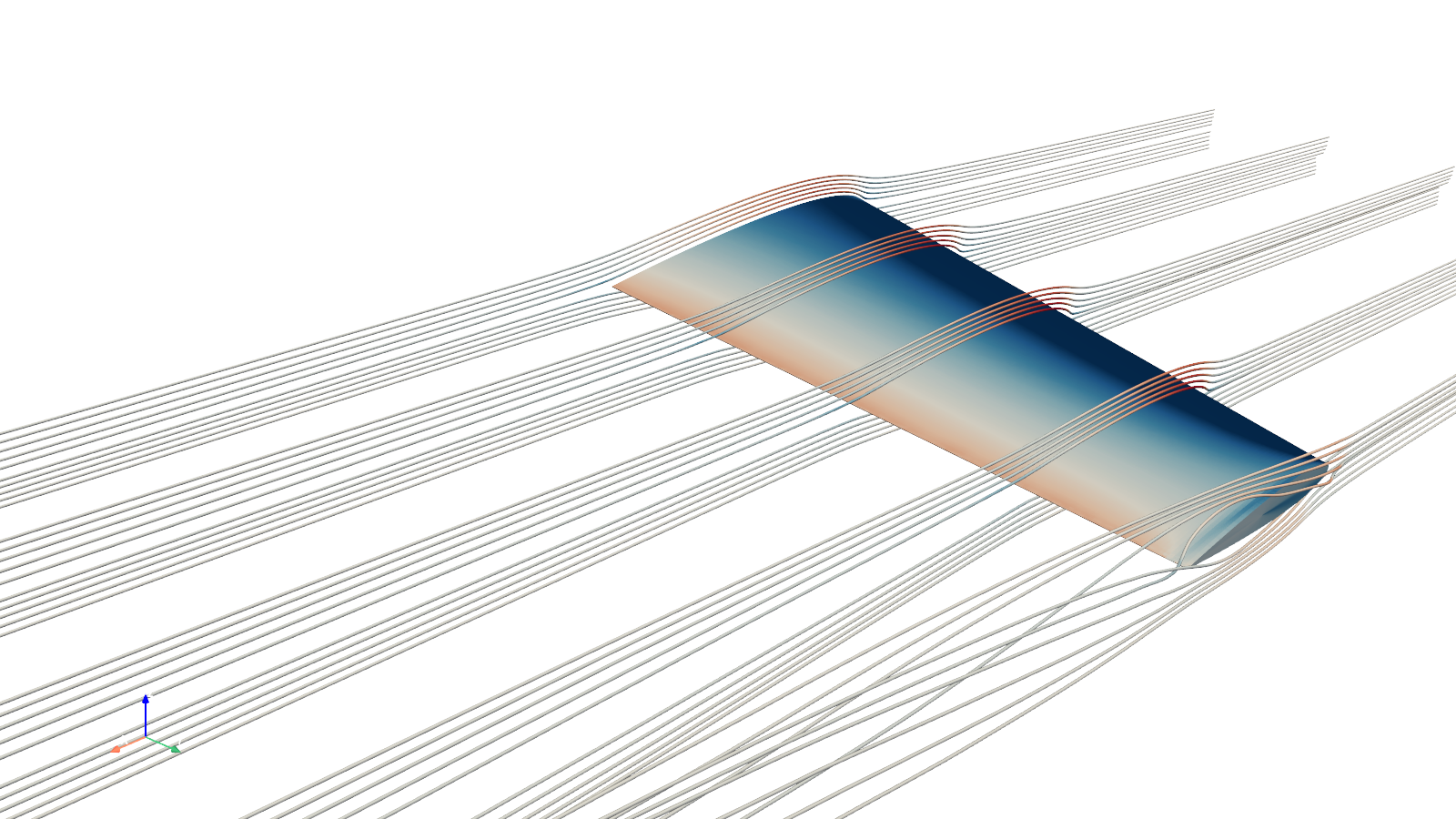}
        \caption{Chord root: 0.7}
    \end{subfigure}
    \hfill
    \begin{subfigure}[b]{0.475\textwidth}
        \centering
        \includegraphics[width=\linewidth, trim={0 100pt 0 80pt}, clip]{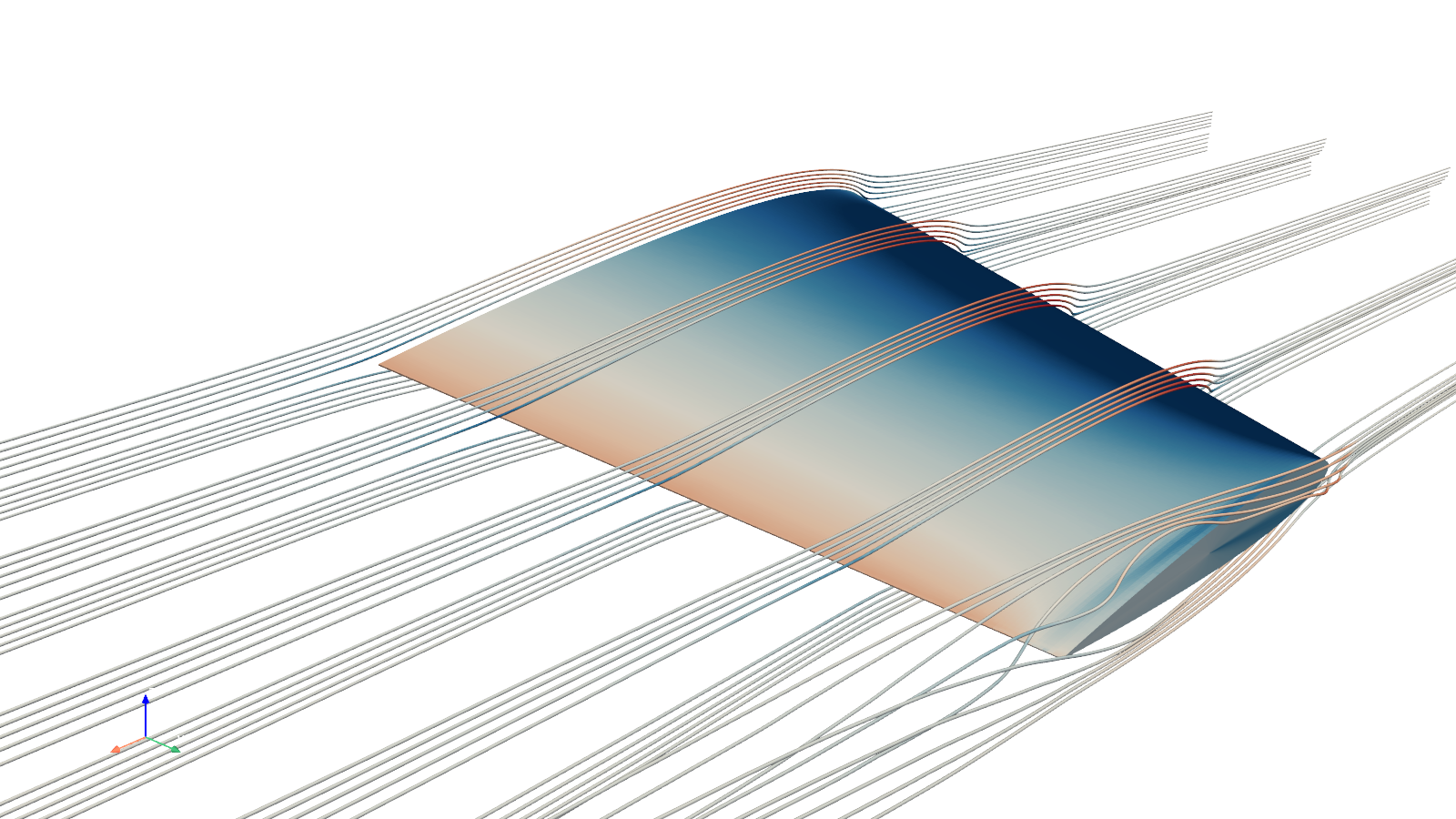}
        \caption{Chord root: 1.2}
    \end{subfigure}

    \begin{subfigure}[b]{0.475\textwidth}
        \centering
        \includegraphics[width=\linewidth, trim={0 100pt 0 80pt}, clip]{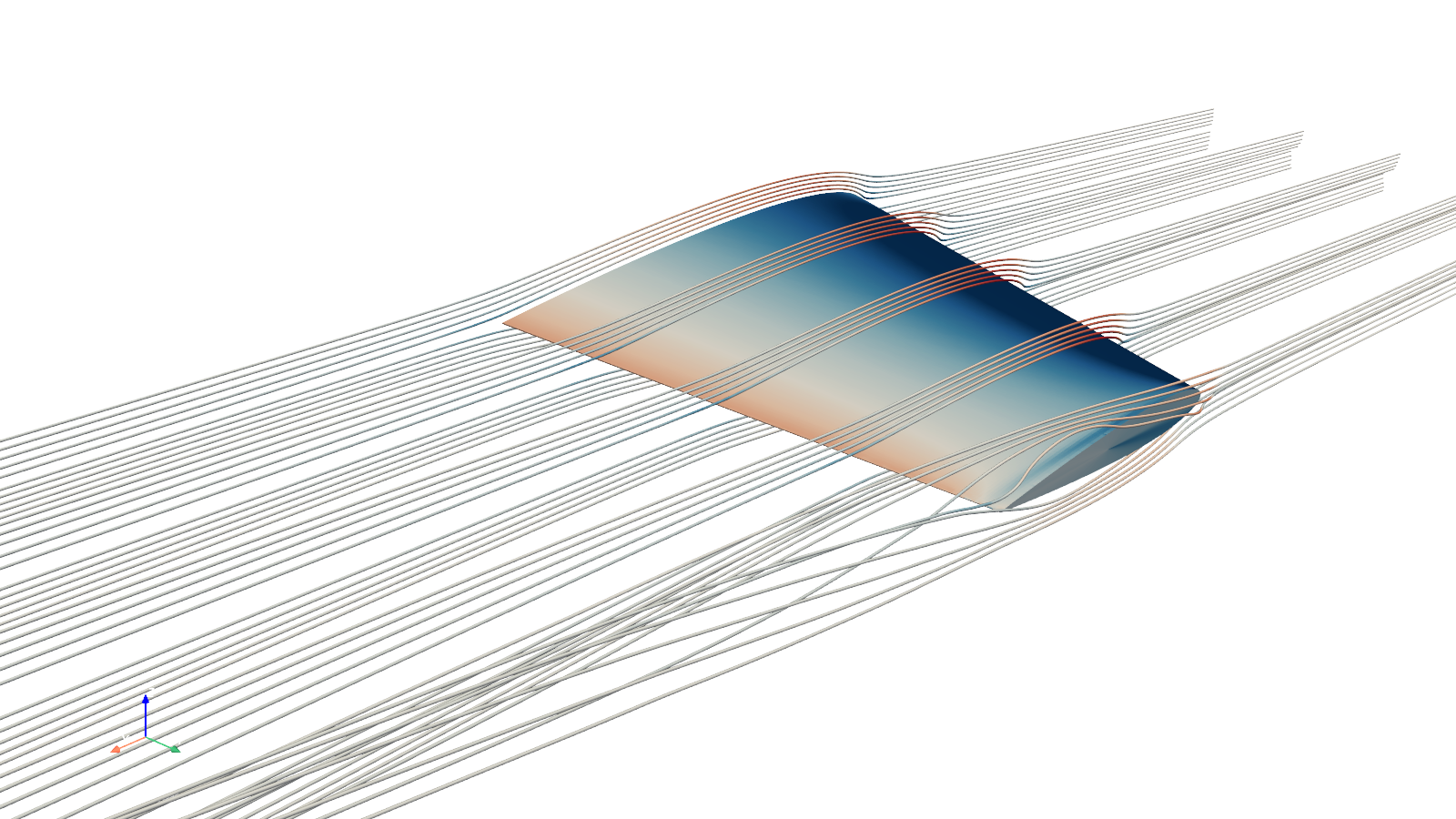}
        \caption{Span: 1.0}
    \end{subfigure}
    \hfill
    \begin{subfigure}[b]{0.475\textwidth}
        \centering
        \includegraphics[width=\linewidth, trim={0 100pt 0 80pt}, clip]{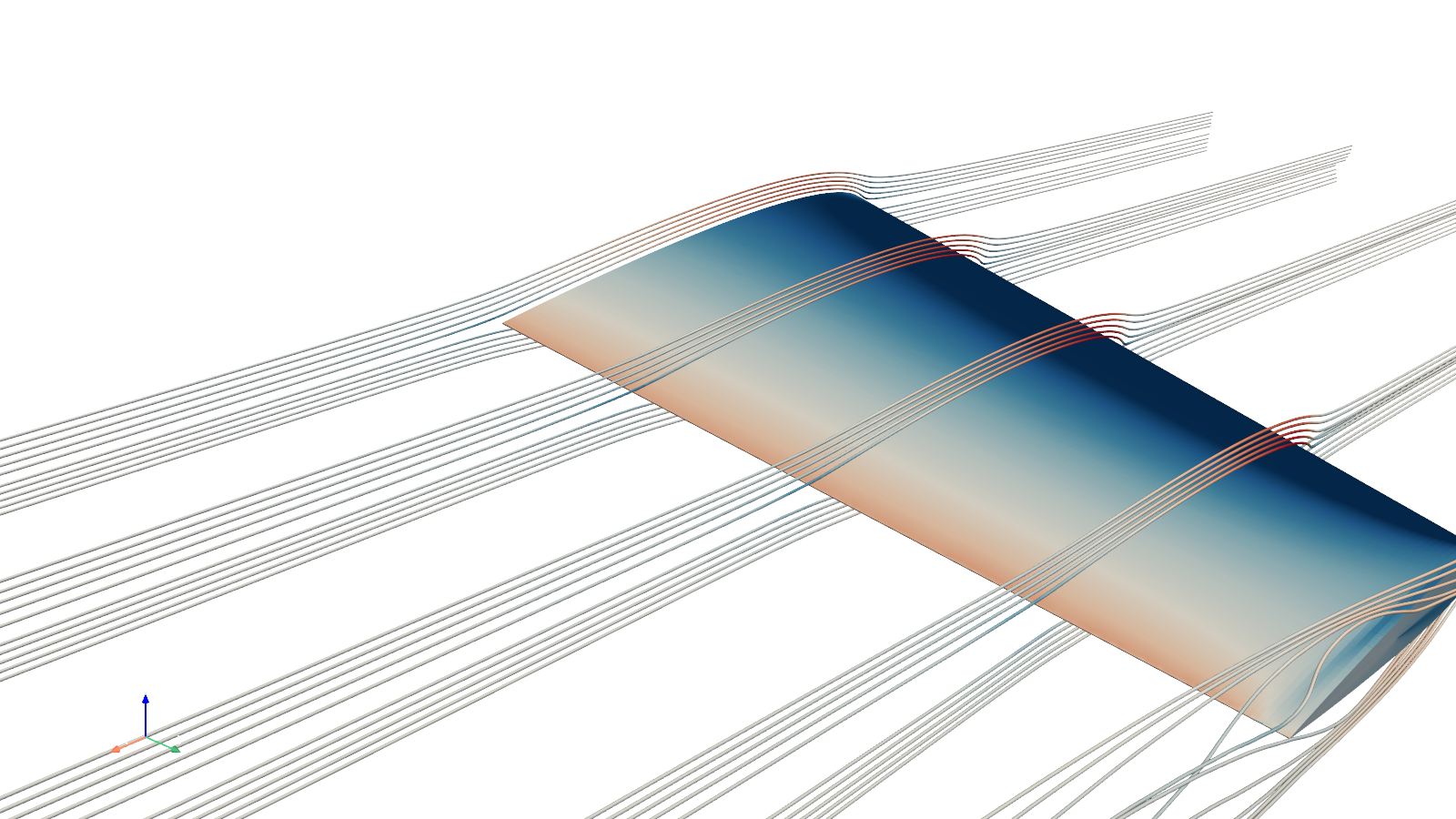}
        \caption{Span: 1.5}
    \end{subfigure}

    \begin{subfigure}[b]{0.475\textwidth}
        \centering
        \includegraphics[width=\linewidth, trim={0 100pt 0 80pt}, clip]{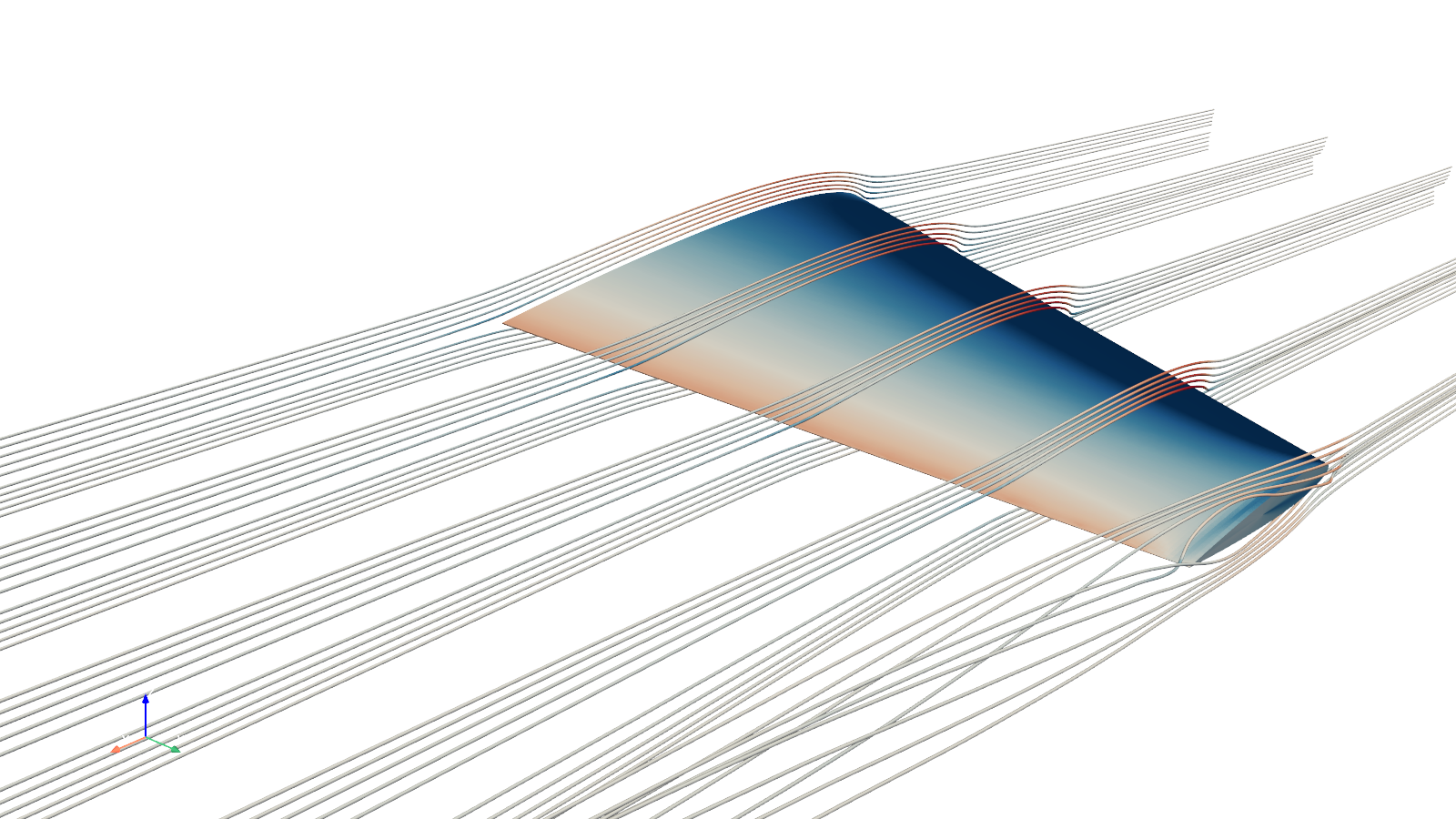}
        \caption{Taper ratio: 0.4}
    \end{subfigure}
    \hfill
    \begin{subfigure}[b]{0.475\textwidth}
        \centering
        \includegraphics[width=\linewidth, trim={0 100pt 0 80pt}, clip]{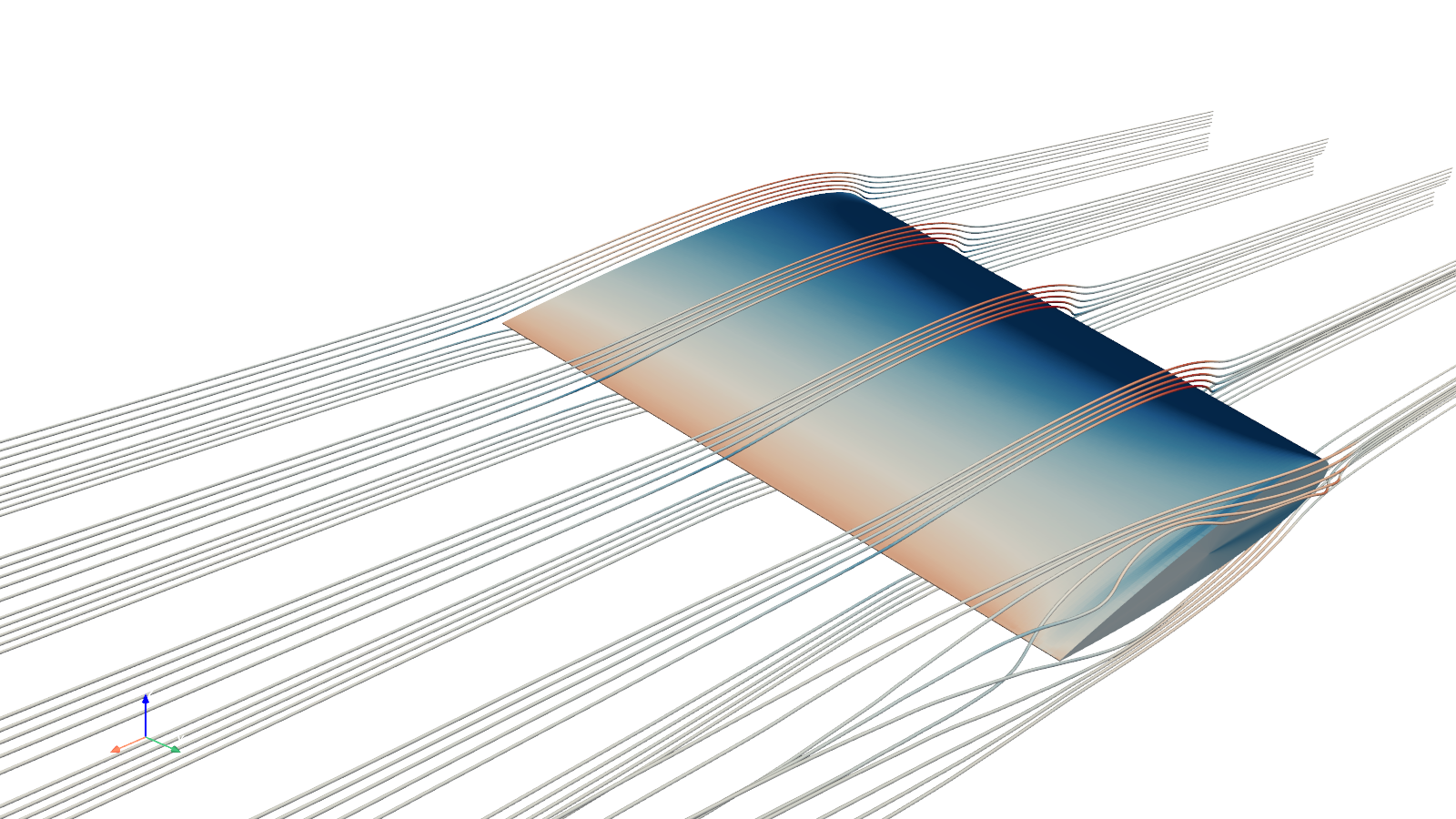}
        \caption{Taper ratio: 0.7}
    \end{subfigure}

    \begin{subfigure}[b]{0.475\textwidth}
        \centering
        \includegraphics[width=\linewidth, trim={0 100pt 0 80pt}, clip]{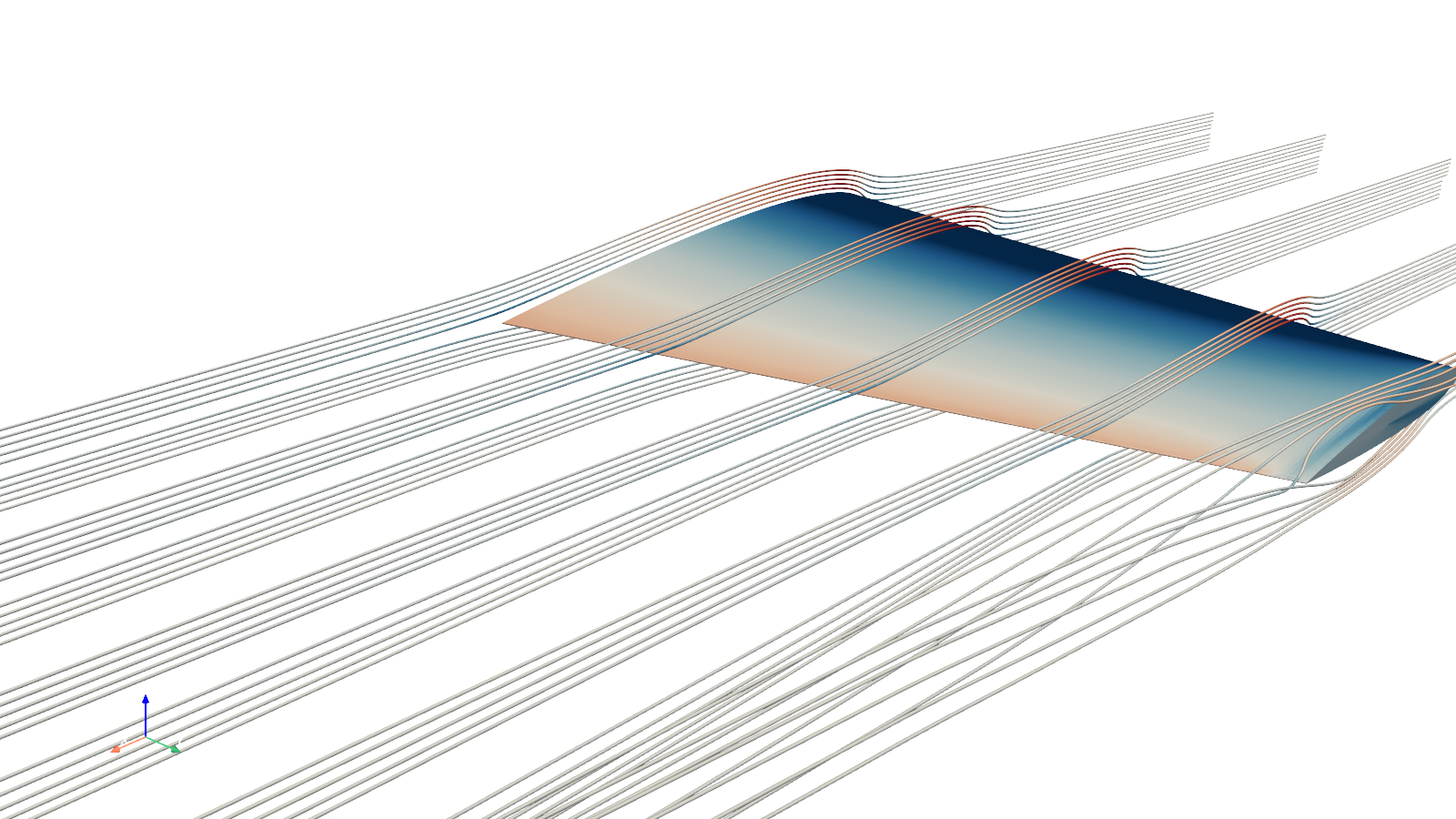}
        \caption{Sweep angle: 0.0}
    \end{subfigure}
    \hfill
    \begin{subfigure}[b]{0.475\textwidth}
        \centering
        \includegraphics[width=\linewidth, trim={0 100pt 0 80pt}, clip]{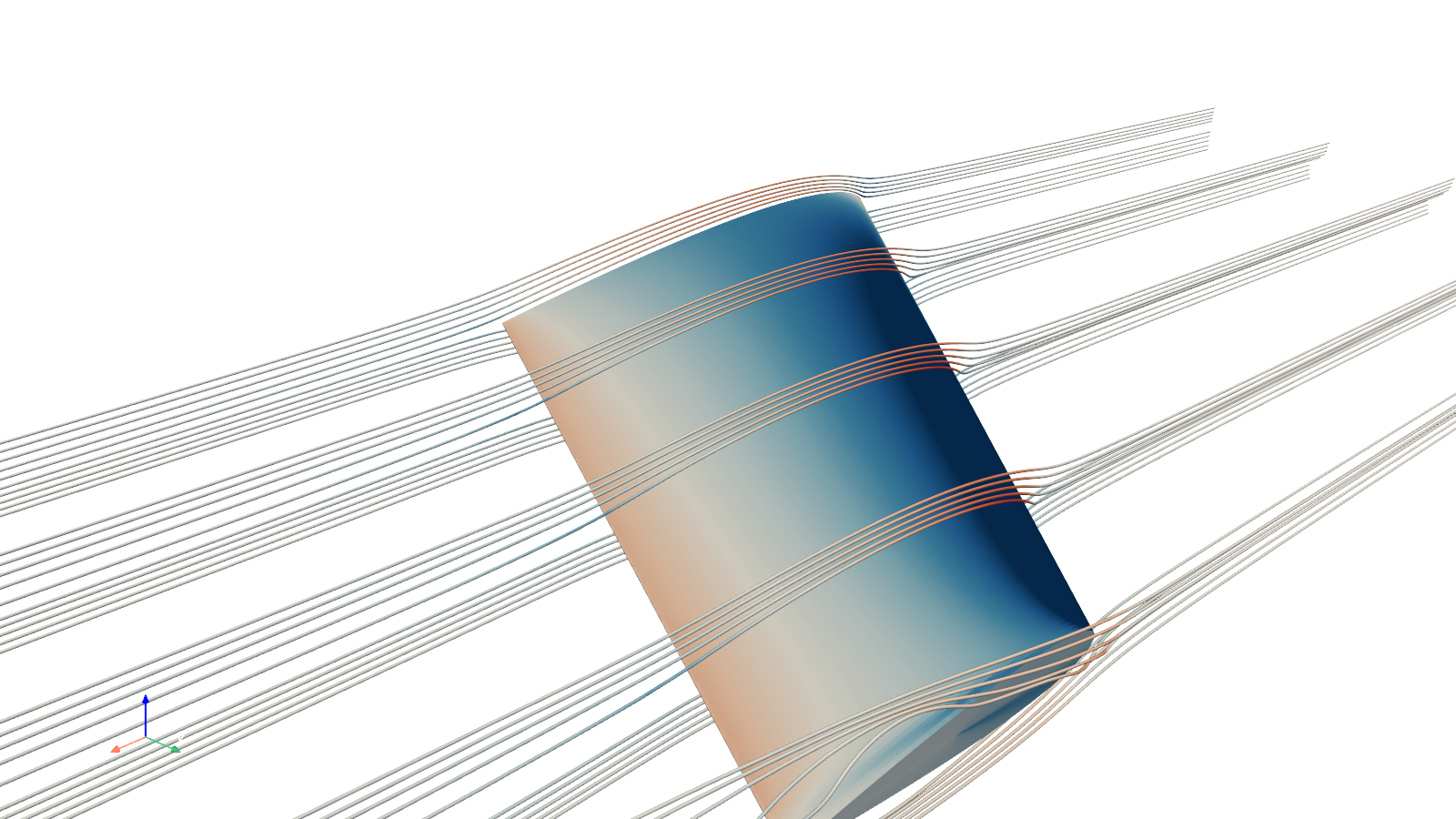}
        \caption{Sweep angle: 40.0}
    \end{subfigure}

    \begin{subfigure}[b]{0.475\textwidth}
        \centering
        \includegraphics[width=\linewidth, trim={0 100pt 0 80pt}, clip]{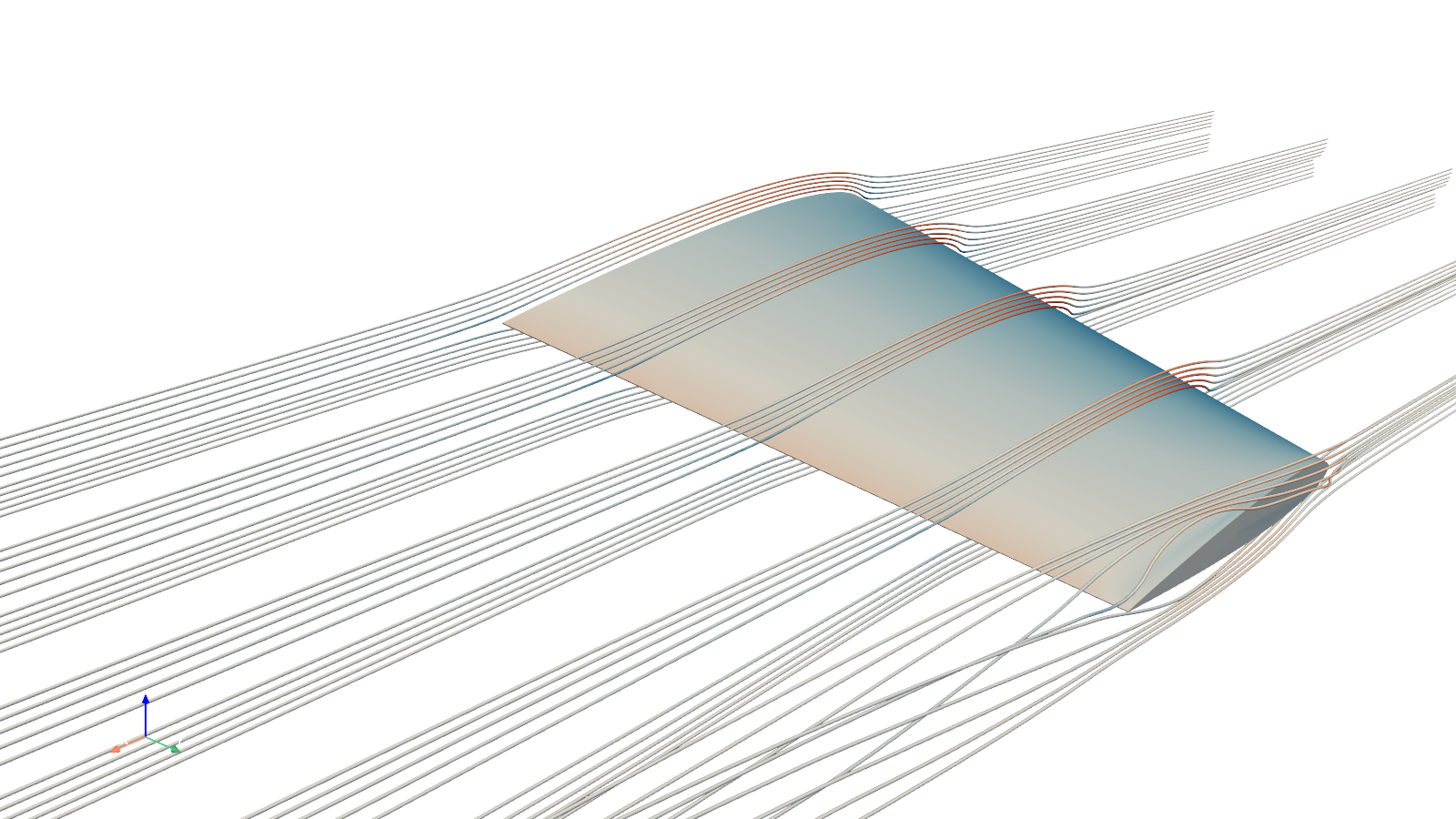}
        \caption{Velocity: 150.0}
    \end{subfigure}
    \hfill
    \begin{subfigure}[b]{0.475\textwidth}
        \centering
        \includegraphics[width=\linewidth, trim={0 100pt 0 80pt}, clip]{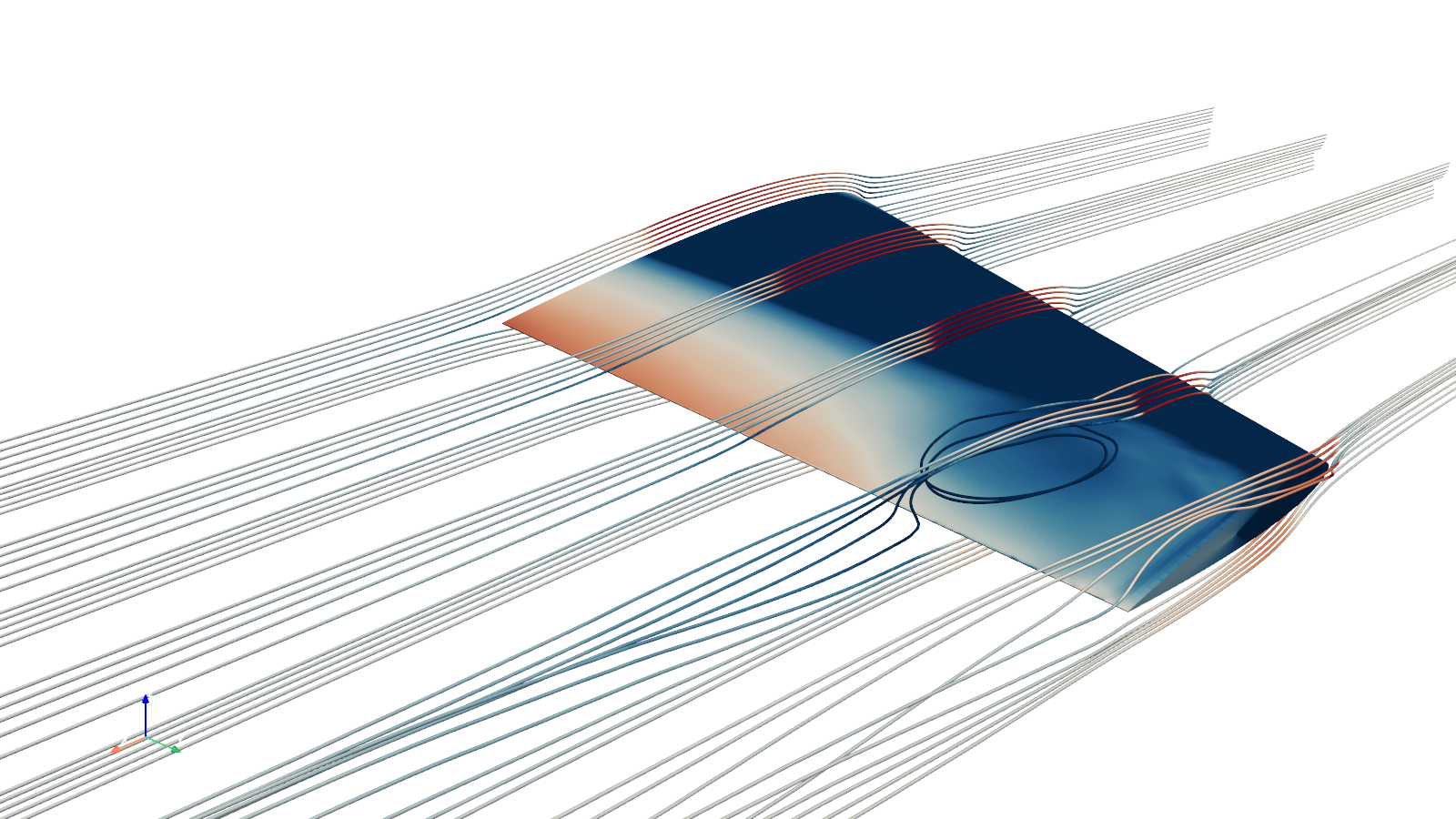}
        \caption{Velocity: 300.0}
    \end{subfigure}

    \begin{subfigure}[b]{0.475\textwidth}
        \centering
        \includegraphics[width=\linewidth, trim={0 100pt 0 80pt}, clip]{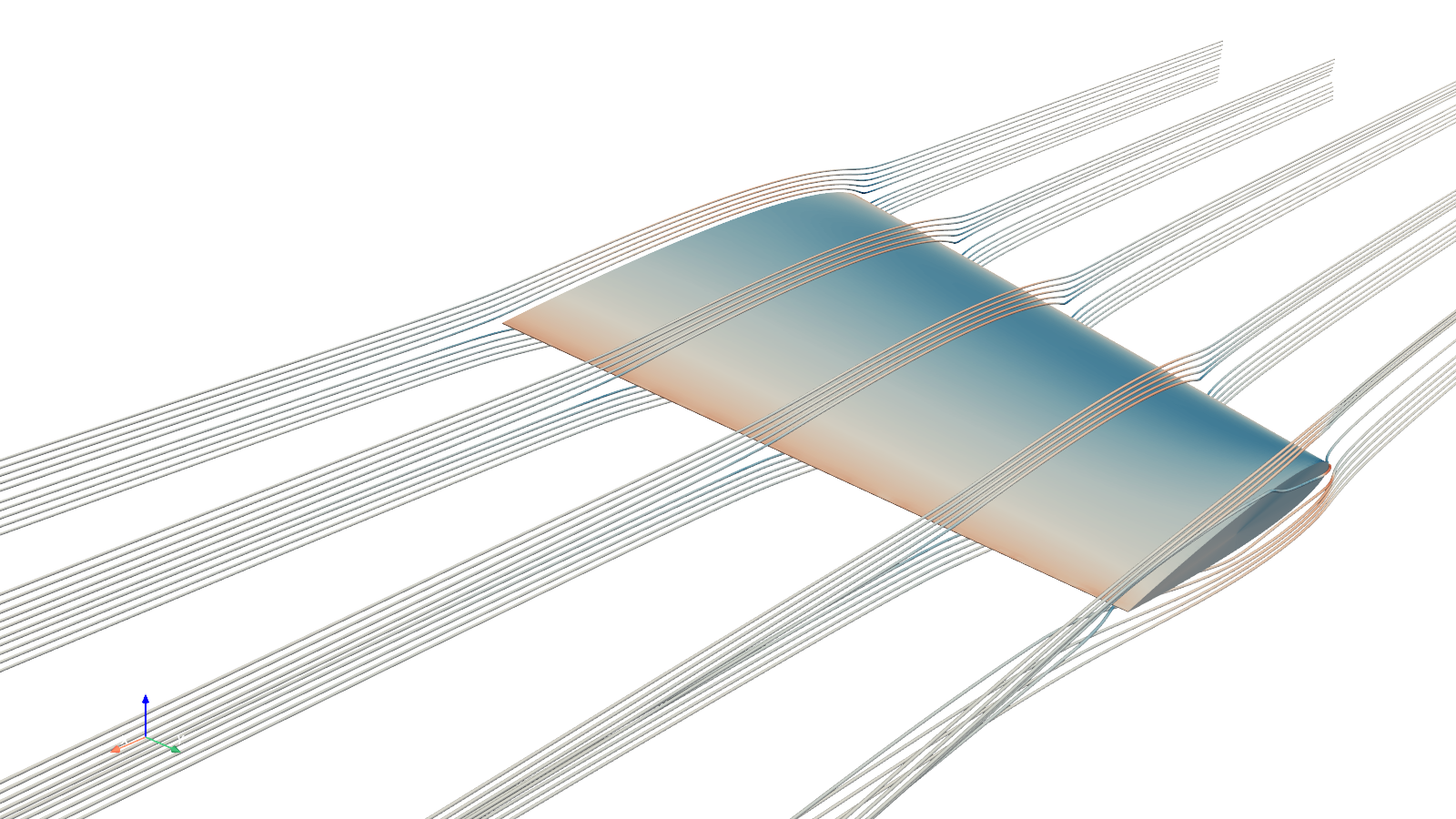}
        \caption{Angle of attack: 0.0}
        \label{aoam2}
    \end{subfigure}
    \hfill
    \begin{subfigure}[b]{0.475\textwidth}
        \centering
        \includegraphics[width=\linewidth, trim={0 100pt 0 80pt}, clip]{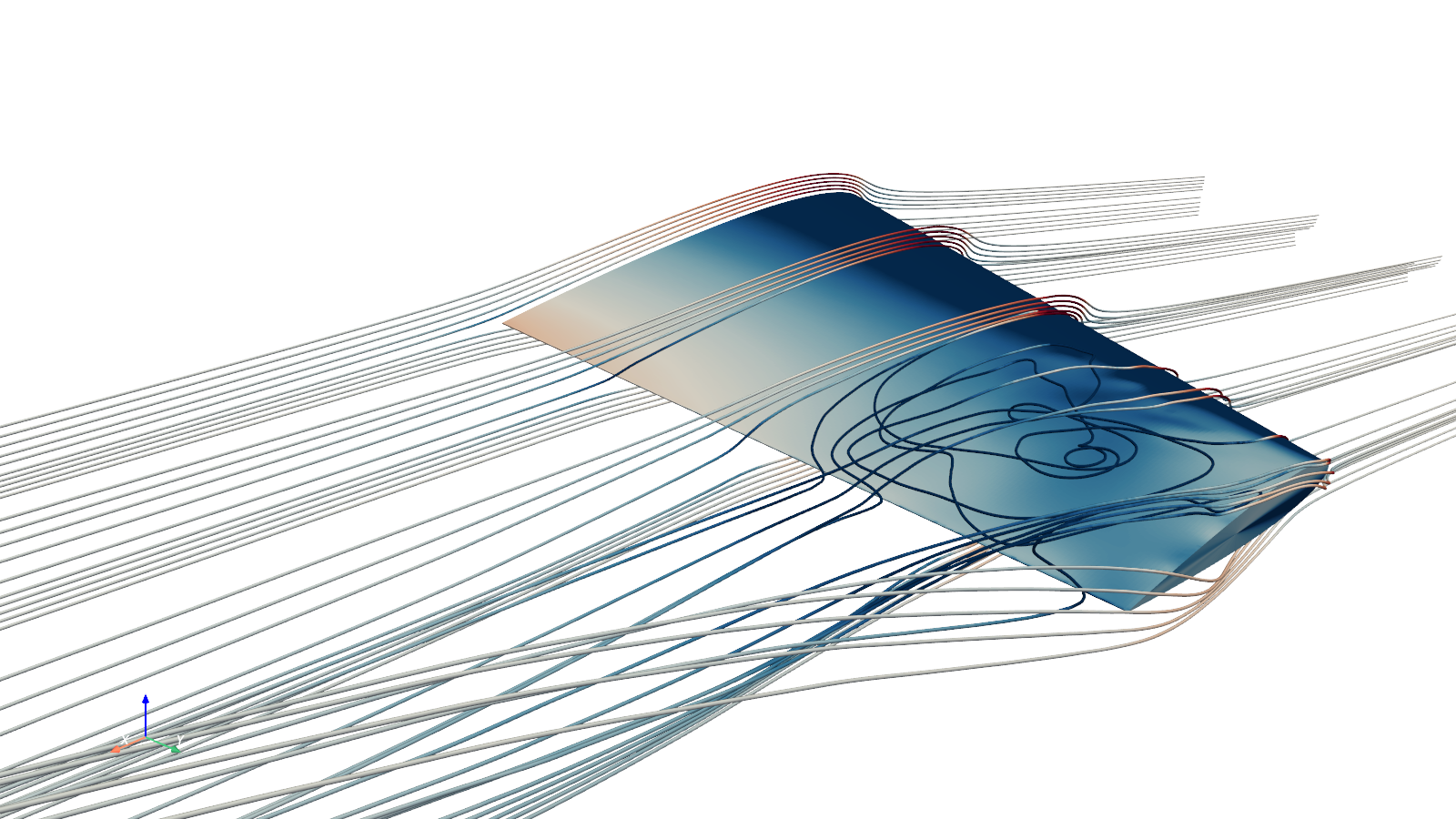}
        \caption{Angle of attack: 10.0}
        \label{aoa10}
    \end{subfigure}

    \caption{Wing geometry with corresponding surface pressure and volume velocity fields for various geometry design parameters and inflow conditions. In each row, the minimum and maximum values for each parameter are visualized, while all other parameters are set to their mean value. Except for \subref{aoam2} and \subref{aoa10}, the angle of attack shown is 4 degrees.}
    \label{fig:parameter_sweep_minmax}
\end{figure}

In Figure \ref{fig:parameter_sweep_minmax}, we show the diversity of the dataset by visualizing the wing geometry and the corresponding surface pressure and volume velocity streamlines for various geometry parameters and inflow conditions. 

\end{document}